%% file: 0_main_ojcoms.tex
\documentclass{IEEEoj}
\usepackage[T1]{fontenc}
\usepackage[utf8]{inputenc}

\usepackage[backend=biber, style=ieee, sorting=none, 
isbn=false, maxnames=99, minnames=99,doi=false]{biblatex}
\usepackage{amsmath,amssymb,amsfonts}
\usepackage{algorithmic}
\usepackage{graphicx,color}
\usepackage{textcomp}
\usepackage{xcolor}
\usepackage{comment}
\usepackage{algorithm,algorithmic}

\newcommand{\V}[1]{\bm{#1}}

\newcommand{\ist}{\hspace*{.3mm}}
\newcommand{\rmv}{\hspace*{-.3mm}}
\newcommand{\iist}{\hspace*{1mm}}

\newcommand{\cmt}[1]{}
\providecommand{\norm}[1]{\lVert#1\rVert}

\def\BibTeX{{\rm B\kern-.05em{\sc i\kern-.025em b}\kern-.08em
T\kern-.1667em\lower.7ex\hbox{E}\kern-.125emX}}
\AtBeginDocument{\definecolor{ojcolor}{cmyk}{0.93,0.59,0.15,0.02}}
\def\OJlogo{\vspace{-4pt}\hskip-4pt\includegraphics[height=18pt]{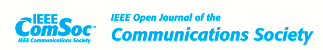}}

\def\authorrefmark#1{\ensuremath{^{\textbf{#1}}}}

\usepackage{soul}
\usepackage{outlines}
\usepackage[thmmarks,amsmath]{ntheorem}
\theoremnumbering{arabic}
\theoremstyle{break}
\theoremsymbol{}
\theoremheaderfont{\normalfont\bfseries}
\theorembodyfont{\upshape}
\theoremstyle{nonumberbreak}
\usepackage{bm}
\usepackage{sansmath}
\usepackage[fleqn]{nccmath}
\usepackage{comment}
\usepackage{multicol}
\usepackage{siunitx}
\usepackage{makecell}

\usepackage[table]{xcolor}
\usepackage{tabularx}
\usepackage{booktabs}
\usepackage{threeparttable}

\usepackage{subfig}
\usepackage[figurename=Figure,tablename=Table]{caption}
\usepackage[noabbrev]{cleveref}

\crefname{figure}{Fig.}{Figs.}
\Crefname{figure}{Fig.}{Figs.}

\crefname{table}{Table}{Tables}
\Crefname{table}{Table}{Tables}

\crefformat{equation}{(#2#1#3)}
\Crefformat{equation}{Equation~(#2#1#3)}

\crefrangeformat{equation}{(#3#1#4)--(#5#2#6)}
\crefmultiformat{equation}{(#2#1#3)}{, (#2#1#3)}{, (#2#1#3)}{, (#2#1#3)}

\usepackage[]{glossaries}
\loadglsentries{abbr}

\begin{document}
%\receiveddate{XX Month, XXXX}
%\reviseddate{XX Month, XXXX}
%\accepteddate{XX Month, XXXX}
%\publisheddate{XX Month, XXXX}
%\currentdate{11 January, 2024}
%\doiinfo{OJCOMS.2024.011100}
\title{Cellular Signal Constructed Convolutional Vision Transformer for High Accuracy Positioning}
\author{Junshi Chen\authorrefmark{1}, Graduate Student Member, IEEE, Xuhong Li\authorrefmark{1}, Member, IEEE, \\ Russ Whiton\authorrefmark{2} and Fredrik Tufvesson\authorrefmark{1}, Fellow, IEEE}
\affil{Department of Electrical and Information Technology, Lund University, Lund, Sweden.}
\affil{European Space Agency, Keplerlaan 1, NL-2200 AG Noordwijk, The Netherlands.}
\corresp{Corresponding author: Junshi Chen (email: junshi.chen@eit.lth.se).}
\authornote{This work has been partially supported by the Vinnova project MIMO-PAD (grant no. 2018-05000), the competence center NextG2Com and the strategic research area ELLIIT''}
\markboth{Cellular Signal Constructed Convolutional Vision Transformer for High Accuracy Positioning}{Chen {et al.}}
\begin{abstract}
Modern cellular systems employ wide bandwidths and large antenna arrays to meet high data rate requirements. The high spatial and temporal resolution for communication also enables high-accuracy positioning as an ancillary benefit. Standard convolutional neural networks (CNNs) and vision Transformers have demonstrated excellent performance in positioning by leveraging delay-angle domain channel representations. However, they still face practical challenges in complicated cellular environments with low signal-to-noise ratios and severe inter-cell interference. This paper proposes a hybrid convolutional vision Transformer (ConViT) architecture that integrates the local receptive fields of CNNs to suppress local noise and employs Transformers to capture global attention among different multipath components. Various fusion strategies for combining signals from multiple distributed base stations are also evaluated. An extended Kalman filter with sensor fusion is applied to further mitigate long tail fluctuations of model estimates. Comprehensive validation is conducted with commercial long-term-evolution signals received by a large antenna array in urban environments with non line-of-sight signals and strong inter-cell interference. ConViT achieves a distance root mean square error (RMSE) of $3.46$\,m and a yaw RMSE of $2.54^{\circ}$, significantly outperforming benchmark models, while maintaining a lower parameter count and reduced computational complexity. Finally, a correspondence analysis between delay-angle power distributions and Transformer attention weights demonstrates the interpretability of the model. 
\end{abstract}
\begin{IEEEkeywords}
Convolutional neural network, Convolutional vision Transformer, Large antenna array, Localization, Machine learning, Positioning, Transformer, Vision Transformer.
\end{IEEEkeywords}
\maketitle
\section{Introduction}\label{sec:intro}
\input{1_Introduction}
\section{Signal Model}\label{sec:channel_model}
\input{2_channel_model}

\section{Cellular Signals Constructed \gls{vit} and \gls{cvit} for Positioning}\label{sec:build_vit}
\input{3_cellavit_structure}
\section{Extended Kalman Filter with Sensor Fusion}\label{sec:ekf}
\input{4_ekf}
\section{Measurement Setup}\label{sec:measurement_set}
\input{5_Measurements_setup}
\section{Measurement Results}\label{sec:measurement_result}
\input{6_measurement_results}

\section{Conclusion}\label{sec:conclusion}
\input{7_conclusion}
\section*{Acknowledgment}\label{sec:ack}
\input{8_acknowledgment}
%\bibliographystyle{IEEEtran}
%\bibliography{IEEEabrv,liberry}
\printbibliography
\end{document}

%% file: abbr.tex
% Comply with https://www.ieee.org/content/dam/ieee-org/ieee/web/org/pubs/ieee-thesaurus.pdf
\newacronym{2d}{2D}{two-dimensional}
\newacronym{2g}{2G}{second generation}
\newacronym{3d}{3D}{three-dimensional}
\newacronym{3gpp}{3GPP}{third generation partnership project}
\newacronym{5g}{5G}{fifth generation}
\newacronym{6g}{6G}{sixth generation}
\newacronym{1pps}{1PPS}{1 pulse-per-second}
\newacronym{acf}{ACF}{autocorrelation function}
\newacronym{acsi}{ACSI}{Absolute Channel State Information}
\newacronym{agc}{AGC}{automatic gain control}
\newacronym{ann}{ANN}{artificial neural network}
\newacronym{aoa}{AOA}{angle-of-arrival}
\newacronym{aod}{AOD}{angle-of-departure}
\newacronym{asic}{ASIC}{application specific integrated circuit}
\newacronym{awgn}{AWGN}{additive white Gaussian noise}
\newacronym{bcsi}{BCSI}{baseline channel state information}
\newacronym{bdir}{BDIR}{beam-domain impulse response}
\newacronym{bp}{BP}{belief propagation}
\newacronym{bs}{BS}{base station}
\newacronym{cbsm}{CBSM}{correlation-based stochastic model}
\newacronym{ccw}{CCW}{counterclockwise}
\newacronym{cdf}{CDF}{cumulative distribution function}
\newacronym{ce}{CE}{channel estimation}
\newacronym{cnn}{CNN}{convolutional neural network}
\newacronym{cp}{CP}{cyclic prefix}
\newacronym{cr}{CR}{coverage region}
\newacronym{crlb}{CRLB}{Cram\'er-Rao lower bound}
\newacronym{crs}{CRS}{cell-specific reference signal}
\newacronym{csi}{CSI}{channel state information}
\newacronym{cw}{CW}{clockwise}
\newacronym{da}{DA}{data association}
\newacronym{dac}{DAC}{digital-to-analog converter}
\newacronym{dc}{DC}{direct current}
\newacronym{dm}{DM}{diffuse multipath}
\newacronym{dmc}{DMC}{dense multipath component}
\newacronym{dmimo}{D-MIMO}{distributed MIMO}
\newacronym{eadf}{EADF}{effective aperture distribution functions}
\newacronym{kf}{KF}{Kalman filter}
\newacronym{ekf}{EKF}{extended Kalman filter}
\newacronym{en}{EN}{energy neutral}
\newacronym{end}{END}{energy neutral device}
\newacronym{epu}{EPU}{edge processing unit}
\newacronym{fa}{FA}{false alarm}
\newacronym{fcnn}{FCNN}{fully-connected neural network}
\newacronym{fdd}{FDD}{frequency division duplexing}
\newacronym{fg}{FG}{factor graph}
\newacronym{fpga}{FPGA}{field-programmable gate array}
\newacronym{ft}{FT}{Fourier transform}
\newacronym{gbsc}{GBSC}{geometry-based deterministic stochastic channel}
\newacronym{gbsm}{GBSM}{geometry-based stochastic model}
\newacronym{gdop}{GDOP}{geometric dilution of precision}
\newacronym{gmf}{GMF}{global map feature}
\newacronym{gnss}{GNSS}{global navigational satellite systems}
\newacronym{gps}{GPS}{global positioning system}
\newacronym{gpst}{GPST}{GPS Time}
\newacronym{gsm}{GSM}{Global System for Mobile Communications}
\newacronym{heu}{HEU}{high epistemic uncertainty}
\newacronym{ic}{IC}{integrated circuit}
\newacronym{id}{ID}{information decoding}
\newacronym{iid}{iid}{independent and identically distributed}
\newacronym{imu}{IMU}{inertial measurement unit}
\newacronym{iot}{IoT}{internet of things}
\newacronym{jcas}{JCAS}{joint communication and sensing}
\newacronym{knn}{kNN}{k-Nearest Neighbor}
\newacronym{leu}{LEU}{low epistemic uncertainty}
\newacronym{lhf}{LHF}{likelihood function}
\newacronym{los}{LoS}{line-of-sight}
\newacronym{lsa}{LSA}{large synthetic array}
\newacronym{lsf}{LSF}{large-scale fading}
\newacronym{lte}{LTE}{long term evolution}
\newacronym{lti}{LTI}{linear time-invariant}
\newacronym{m2m}{M2M}{machine-type communication}
\newacronym{mb}{MB}{multi-Bernoulli}
\newacronym{meo}{MEO}{medium earth orbit}
\newacronym{mf}{MF}{map feature}
\newacronym{mfad}{MFAD}{matched filter azimuthal-delay}
\newacronym{mfpd}{MFPD}{matched filter polarimetric-delay}
\newacronym{mimo}{MIMO}{multiple-input multiple-output}
\newacronym{miso}{MISO}{multiple-input single-output}
\newacronym{mmse}{MMSE}{minimum mean-square error}
\newacronym{mmwave}{mmWave}{millimeter wave}
\newacronym{mospa}{MOSPA}{mean \gls{ospa}}
\newacronym{mp}{MP}{multipath-based}
\newacronym{mpc}{MPC}{multipath component}
\newacronym{mpslam}{MP-SLAM}{multipath-based simultaneous localization and mapping}
\newacronym{mrc}{MRC}{maximum ratio combining}
\newacronym{nb}{NB}{narrowband}
\newacronym{nlp}{NLP}{natural language processing}
\newacronym{nlos}{NLoS}{non-line-of-sight}
\newacronym{ofdm}{OFDM}{orthogonal frequency-division multiplexing}
\newacronym{olos}{OLoS}{obstructed-line-of-sight}
\newacronym{ospa}{OSPA}{optimal subpattern assignment}
\newacronym{ota}{OTA}{over-the-air}
\newacronym{pa}{PA}{physical anchor}
\newacronym{pc}{PC}{pilot count}
\newacronym{pdf}{PDF}{probability density function}
\newacronym{pdp}{PDP}{power delay profile}
\newacronym{pf}{PF}{potential feature}
\newacronym{phd}{PHD}{probability hypothesis density}
\newacronym{pl}{PL}{path loss}
\newacronym{ple}{PLE}{pathloss exponent}
\newacronym{pmf}{PMF}{probability mass function}
\newacronym{ppp}{PPP}{Poisson point process}
\newacronym{pr}{PR}{physical reflector}
\newacronym{pss}{PSS}{primary synchronization signal}
\newacronym{ra}{RA}{reflectarray}
\newacronym{rb}{Rb}{rubidium}
\newacronym{rcs}{RCS}{radar cross section}
\newacronym{rf}{RF}{radio frequency}
\newacronym{rfid}{RFID}{radio frequency identification}
\newacronym{rfs}{RFS}{random finite set}
\newacronym{ris}{RIS}{reconfigurable intelligent surfaces}
\newacronym{rms}{RMS}{root mean square}
\newacronym{rmse}{RMSE}{root mean square error}
\newacronym{rnn}{RNN}{recurrent neural network}
\newacronym{roi}{RoI}{region-of-interest}
\newacronym{rs}{RS}{reference signal}
\newacronym{rss}{RSS}{Received Signal Strength}
\newacronym{rssi}{RSSI}{received signal strength indicator}
\newacronym{rtk}{RTK}{real-time kinematic}
\newacronym{rw}{RW}{RadioWeaves}
\newacronym{rx}{Rx}{receiver}
\newacronym{s-parameters}{S-parameters}{scattering parameters}
\newacronym{sa}{SA}{synchronization anchor}
\newacronym{sar}{SAR}{specific absorption rate}
\newacronym{sbl}{SBL}{sparse Bayesian learning}
\newacronym{sdr}{SDR}{software-defined radio}
\newacronym{simo}{SIMO}{single-input multiple-output}
\newacronym{sinr}{SINR}{signal-to-interference-plus-noise ratio}
\newacronym{siso}{SISO}{single-input single-output}
\newacronym{slam}{SLAM}{simultaneous localization and mapping}
\newacronym{smc}{SMC}{specular multipath component}
\newacronym{snr}{SNR}{signal-to-noise ratio}
\newacronym{sp}{SP}{specular path}
\newacronym{spa}{SPA}{sum-product algorithm}
\newacronym{ssf}{SSF}{small-scale fading}
\newacronym{sss}{SSS}{secondary synchronization signal}
\newacronym{std}{STD}{standard deviation}
\newacronym{suca}{SUCA}{stacked uniform circular array}
\newacronym{tdma}{TDMA}{time-division multiple access}
\newacronym{tdoa}{TDOA}{time-difference-of-arrival}
\newacronym{toa}{TOA}{time-of-arrival}
\newacronym{tosm}{TOSM}{through-open-short-match}
\newacronym{tx}{Tx}{transmitter}
\newacronym{ue}{UE}{user equipment}
\newacronym{uhf}{UHF}{ultra high frequency}
\newacronym{ula}{ULA}{uniform linear array}
\newacronym{upa}{UPA}{uniform planar array}
\newacronym{ura}{URA}{uniform rectangular array}
\newacronym{urllc}{URLLC}{ultra-reliable and low-latency communication}
\newacronym{usrp}{USRP}{universal software radio peripheral}
\newacronym{uwb}{UWB}{ultrawideband}
\newacronym{v2x}{V2x}{vehicle-to-everything}
\newacronym{va}{VA}{virtual anchor}
\newacronym{vhf}{VHF}{very high frequency}
\newacronym{vit}{ViT}{vision Transformer}
\newacronym{vivit}{ViViT}{video vision Transformer}
\newacronym{vlf}{VLF}{very low frequency}
\newacronym{vna}{VNA}{vector network analyzer}
\newacronym{vr}{VR}{visibility region}
\newacronym{wb}{WB}{wideband}
\newacronym{wpt}{WPT}{wireless power transfer}
\newacronym{zf}{ZF}{zero forcing}
\newacronym{msa}{MSA}{multiheaded self attention}
\newacronym{mlp}{MLP}{multi-layer perception}
\newacronym{ln}{LN}{layer normalization}
\newacronym{ffn}{FFN}{feed forward network}
\newacronym{nn}{NN}{neural network}
\newacronym{dnn}{DNN}{deep neural network}
\newacronym{ml}{ML}{machine learning}
\newacronym{cvit}{ConViT}{Convolutional ViT}
\newacronym{fdn}{FDN}{feature decode network} 
\newacronym{resnet}{ResNet}{residual network} 
\newacronym{cir}{CIR}{channel impulse response} 
\newacronym{gflop}{GFLOP}{Giga-floating-point operation}
\newacronym{gelu}{GELU}{Gaussian error linear unit}
\newacronym{ici}{ICI}{inter-cell interference}
\newacronym{icic}{ICIC}{inter-cell interference cancellation}

%% file: 1_Introduction.tex
Beyond \gls{5g} cellular networks are expected to deliver both extreme wireless connectivity and high-accuracy positioning and sensing capabilities \cite{harsh_6g_paper, 1g_to_5g_cellular_localization}. This evolution is enabled by large-scale antenna arrays, wide bandwidths, and dense and distributed network architectures, which jointly provide higher spatial and temporal resolution and improved signal availability \cite{christian2024dmimo, Benjamin2024dmimo}. These advances make cellular-based positioning promising for supporting accuracy-critical services such as the Industrial Internet of Things, extended reality, and intelligent transportation systems \cite{Saleh2026vehicularpositioning, whiton2022cellular,Ko2021v2xpos}. It can also serve as a vital complement to \gls{gnss} \cite{dwivedi2021positioning, kassas_magazine,Brambilla2024gnss5g}, particularly in urban canyons and indoor environments, where \gls{gnss} signals are often degraded by multipath propagation, severe attenuation, or blockage \cite{zhu2018gnss,GrovesPrinciples}.

Cellular based positioning methods generally fall into two categories. Classical model-based methods extract geometric information, such as the \gls{toa} and \gls{aoa} of propagation paths, from received signals for triangulation and trilateration \cite{Yang2022ltetriang}. However, these approaches are sensitive to estimation biases caused by \glspl{mpc} reflected or scattered by environmental features. To mitigate this issue, multipath-based \gls{slam} methods have been extensively investigated. These methods exploit \glspl{mpc} by associating them with environmental features, such as \glspl{va} \cite{ChSLAMOrig,LiLeiCaiTuf:ICC2024} for specular reflections, point scatterers \cite{Kim2024TSP} and reflective surfaces \cite{deutschmann2026TSP, LiTWC2026,LiFUSION2026}, and jointly estimate the mobile agent state and the feature states. Although effective, \gls{slam} methods rely on explicit modeling of environmental features and system responses, including hardware imperfections. This can lead to model mismatch in practical deployments \cite{1g_to_5g_cellular_localization}. To overcome these limitations, data-driven \gls{ml} methods \cite{venus2026FUSION,pan2026aipossurvey, Sze2017dnnpossurvey} have been investigated for positioning. These methods estimate positions directly from raw signals or their transformed representations in the frequency, delay, and angular domains, demonstrating robustness to noise and environmental fluctuations. Delay-angle representations are usually preferred for positioning since they reflect the physical geometries of cellular channels \cite{russ2024wiometric}. 

Two mainstream \gls{dnn} models have been developed with different focuses. \Gls{cnn} based positioning methods \cite{krizhevsky2017imagenet, guoda2024fcnn} have demonstrated strong performance by treating delay-angle representations analogously to images \cite{joaovieira2017deep,russ2024wiometric, sun2019fingerprint}. However, \gls{cnn} models mainly focus on local \gls{mpc} representations. In contrast, attention-based Transformers \cite{Vaswani2017attention,guoda2024attention,ilayda20255gattention}, particularly \gls{vit} \cite{dosovitskiy2021vit,Hyung2025vit, Gong2024vit} architectures, which are suitable for delay-angle representations, capture global dependencies across different \gls{mpc} representations and demonstrate better performance. 

Despite their superior performance, applying \gls{vit} architectures to positioning faces several practical challenges. First, \glspl{vit} typically require a large amount of training data to outperform their \gls{cnn} counterparts \cite{wu2021cvt}, whereas collecting accurately labeled field measurement data is costly and time-consuming. Second, cellular signals in the delay-angle domain exhibit image-like local structures, with strong correlations among spatially adjacent representations. A standard \gls{vit} architecture does not explicitly exploit this local correlation. Third, the resolution of the delay-angle representation is usually constrained by the signal bandwidth and the array aperture, making weak \gls{mpc} features difficult to distinguish from sidelobes and noise. 

\Gls{ici} introduces an additional challenge. Two adjacent \glspl{bs} may transmit downlink signals simultaneously, causing \gls{ici} at the mobile agent and degrading signal quality and positioning performance. \Gls{icic} algorithms \cite{lte_ic,rimax_ic} can separate the received signals from these two \glspl{bs}, enabling their joint use to enhance positioning performance \cite{Khalife2019dop,Li2020dop, langley1999dilution,junshi2026gmf}. Such multi-\gls{bs} processing strategies may also be extended to beyond \gls{5g} systems employing distributed \gls{mimo} \cite{christian2024dmimo}. However, the separated signal powers remain imbalanced, and residual interference persists in some locations. These nonideal effects distort the delay-angle structures and may bias the attention mechanism toward interference components rather than weak but geometrically informative \gls{mpc} features. Therefore, directly applying a conventional \gls{vit} to the separated signals may not guarantee robust performance.

To address these limitations, we adopt a \gls{cvit}-based architecture \cite{wu2021cvt,lin2023ConvFormer,gu2023conformer,d2021convit,guo2022cmt}, whose application to cellular positioning in complicated environments remains limited. First, the delay-angle representations of the received signals are processed by a \gls{cnn} backbone to suppress noise and residual interference, thereby effectively extracting reliable local features. Second, the features are fed into a Transformer to capture global dependencies and ensure accurate positioning. Finally, the model estimates are filtered using an \gls{ekf} with sensor fusion to further suppress sudden fluctuations. 

We further investigate different strategies for integrating signals from multiple \glspl{bs} under power imbalance and residual interference. Although the Transformer architecture can readily integrate signals from two \glspl{bs}, the optimal fusion strategy under these nonideal conditions remains insufficiently explored. A straightforward method is to directly concatenate the signals and feed them into a single unified model. However, this may result in cross attention over noisy signals and lead to biased optimization in certain dimensions. In this paper, we evaluate different combination strategies and propose a \gls{cvit} architecture supporting two \glspl{bs}. It processes the signal from each \gls{bs} independently using a dedicated \gls{cnn} and encoder and performs cross attention between the two \glspl{bs} at a later stage of the Transformer. This architecture ensures that robust and \gls{bs}-specific features are extracted before fusion, resulting in more balanced and resilient positioning performance. 

The proposed \gls{cvit} model is validated using field measurement data collected from a commercial \gls{lte} network with a large antenna array in the city center of Lund, Sweden. The measurements include nonidealities such as interference, \gls{nlos} propagation, multipath effects, as well as information from multiple \glspl{bs}, making them highly relevant to beyond \gls{5g} systems. The proposed model is benchmarked against a representative \gls{cnn} (\gls{resnet} \cite{he2016deep}) and several \gls{vit} architectures. The results demonstrate that, even with a limited amount of training data, \gls{cvit} achieves superior positioning performance while maintaining a smaller parameter count and lower computational complexity than the reference models. Additionally, a comparison between the input signal power distribution in the delay-angle domain and the Transformer attention weights demonstrates the interpretability of the model. 

In summary, we present a \gls{cvit} model that is suitable for complicated cellular environments with \gls{nlos} propagation and \gls{ici}, and explore several critical aspects of the architecture:
\begin{itemize} 
\item We apply the \gls{cvit} architecture to the delay-angle domain of cellular signals to exploit the geometric characteristics of wireless channels. It leverages \glspl{cnn} for local feature refinement and Transformers for global dependency modeling.
\item We investigate different multi-\gls{bs} fusion strategies and propose \gls{bs}-specific encoders with late cross-attention to mitigate power imbalance and residual interference. 
\item We perform a comprehensive benchmark of different models using real-world measurement data, providing a detailed analysis of computational complexity and joint 2D position $(x, y)$ and yaw estimation performance, the latter of which remains underexplored in \gls{ml}-based cellular positioning. 
\item We validate the interpretability of the model by demonstrating the correspondence between the delay-angle power distribution and the Transformer's attention weights. 
\end{itemize}

The remainder of this paper is organized as follows. \Cref{sec:channel_model} introduces the signal model and the delay-angle representation of the signal. \Cref{sec:build_vit} details the construction of the \gls{vit} and \gls{cvit} models in the delay-angle domain. \Cref{sec:ekf} introduces the \gls{ekf} framework with sensor fusion to filter the model outputs. \Cref{sec:measurement_set} describes the measurement system setup, followed by \Cref{sec:measurement_result}, which provides a comprehensive analysis of the model parameters, computational complexity, and positioning performance. Finally, \Cref{sec:conclusion} concludes the paper. 

\textit{Notation}: Matrices and vectors are respectively denoted by uppercase and lowercase boldface letters, e.g., $\mathbf{A}$ and $\bm{a}$. The superscripts $ (\cdot)^{\mathrm{T}}$ and $(\cdot)^{\mathrm{H}}$ denote the matrix transpose and Hermitian transpose. The operators $\odot$ and $\diamond$ represent the Hadamard and Khatri–Rao products. $\|\cdot\|_{\mathrm{F}}$ represents the  Frobenius norm. The speed of light is $ c \simeq 3 \cdot 10^8$ m/s.

%% file: 2_channel_model.tex
We consider a cellular system in a \gls{3d} scenario with horizontal and vertical signal propagation. There are $I$ \glspl{pa} transmitting radio signals at fixed positions $\V{p}_{\mathrm{pa}}^i = [p_{\mathrm{pa},\mathrm{x}}^i \iist p_{\mathrm{pa},\mathrm{y}}^i \iist p_{\mathrm{pa},\mathrm{z}}^i ]^{\mathrm{T}}$, $i \in \{1,\dots,I\}$. At each discrete time step $ n $, a mobile agent receives the signals at position $\V{p}_n=[p_{\mathrm{x},n}\, p_{\mathrm{y},n}\,p_{\mathrm{z},n}]^\mathrm{T} $ with orientation, i.e., yaw, pitch, and roll, $\bm{o}_n = [\psi_{n}, \theta_{n}, \phi_n]^{\mathrm{T}}$, both the position and orientation are unknown and time-varying. Each \gls{pa} and the mobile agent constitute a \gls{simo} system, where each \gls{pa} is equipped with a single dual-polarized antenna, while the mobile agent is equipped with a switched antenna array of $M$ ports. The transmitted signals from the \glspl{pa} interact with objects in the environment, resulting in \glspl{mpc} received by the mobile agent. The specularly reflected \glspl{mpc} can be viewed as signals transmitted from \glspl{va}, which are the mirrored positions of the \glspl{pa} with respect to planar surfaces. The \glspl{pa} and \glspl{va} are collectively termed \glspl{mf} with positions $ \V{p}_{l}^i =[p_{l,\mathrm{x}}^i \iist p_{l,\mathrm{y}}^i \iist  p_{l,\mathrm{z}}^i]^{\mathrm{T}}$, where $l \in \{1,\dots,L_{n}^i\}$ and $ L_{n}^i $ denotes the number of visible \glspl{mf} at time step $n$. The \gls{mpc} associated with the $l$th \gls{mf} $\V{p}_{l}^i$ can be characterized by its propagation parameters, including delay $\tau_{l,n}^i$, azimuth \gls{aoa} $\varphi_{l,n}^i$, elevation \gls{aoa} $\theta_{l,n}^i$, and Doppler shift $\nu_{l,n}^i$:
\begin{align}
    \tau_{l,n}^i &= \frac{\|\V{p}_{n} - \V{p}_{l}^i\|}{c} + \Delta\tau_{\mathrm{o}}^i ,  \, \\
    \varphi_{l,n}^i &= \mathrm{atan2}\big(p_{l,{\mathrm{y}}}^i - p_{\mathrm{y},n},\ist p_{l,{\mathrm{x}}}^i - p_{\mathrm{x},n}\big) - \psi_{n}, \, \\
    \theta_{l,n}^i &= \mathrm{asin}\big((p_{l,{\mathrm{z}}}^i - p_{\mathrm{z},n})/\norm{\V{p}_l^i-\V{p}_n}\big) - \theta_{n}  ,
\end{align}
\begin{equation}
    \nu_{l,n}^i = \frac{f_{\mathrm{c}}\V{v}_{n}(\V{p}_l^i-\V{p}_n)^\mathrm{T}}{c\ist\norm{\V{p}_l^i-\V{p}_n}}.
\end{equation}
Here $\Delta\tau_{\mathrm{o}}^i$ is the clock offset between the $i$-th \gls{pa} and the agent, $f_{\mathrm{c}}$ is the carrier frequency, and $\V{v}_{n}$ is the velocity of the mobile agent. 

The received signal at time step $n$ is the sum of \gls{sp} multipaths, \gls{dmc} and noise from all \glspl{pa},  and can be expressed as
\begin{equation}
    \bm{y}(n)=\sum_{i=1}^I \left( {\bm{s}(\boldsymbol{\theta}_{\mathrm{sp},n}^i)}+{\bm{s}_{\mathrm{dmc},n}^i} \right)\odot \bm{x}^i_n +\bm{n}_0 \in \mathbb{C}^{MN_{\mathrm{f}}\times 1}.
\end{equation}
Here, $\bm{x}_n^i$ is the reference signal from the $i$-th \gls{bs}, $\bm{n}_0$ is \gls{awgn}, $M$ is the number of antenna ports, and $N_{\mathrm{f}}$ is the number of subcarriers. The \gls{sp} $\bm{s}(\boldsymbol{\theta}_{\mathrm{sp},n}^i)$ and \gls{dmc} $\bm{s}_{\mathrm{dmc},n}^i$ are elaborated later. Since the received signal contains signals from multiple \glspl{pa}, it is necessary to separate them using \gls{icic} algorithms for each antenna element, such as those in \cite{lte_ic, rimax_ic}. After iterative interference cancellation, the signal from the $i$th \gls{bs} is separated and reshaped into a matrix as: 
\begin{equation}
\mathbf{Y}^i(n)=  {\mathbf{S}(\boldsymbol{\theta}_{\mathrm{sp},n}^i)}+{\mathbf{S}_{\mathrm{dmc},n}^i} \in \mathbb{C}^{M\times N_{\mathrm{f}}}.
\end{equation}
The \gls{sp} response $\mathbf{S}(\boldsymbol{\theta}_{\mathrm{sp},n}^i)$ is given by
\begin{equation}
    \mathbf{S}(\boldsymbol{\theta}_{\mathrm{sp},n}^i)  
    = \mathbf{B}_{\mathrm{rh},n}^i  \boldsymbol{\Gamma}_{\mathrm{h},n}^i  \mathbf{B}^{i{\mathrm{T}}}_{\mathrm{f},n}  + \mathbf{B}_{\mathrm{rv},n}^i \boldsymbol{\Gamma}_{\mathrm{v},n}^i\mathbf{B}^{i{\mathrm{T}}}_{\mathrm{f},n}
\end{equation}
where the basis matrices are defined as
%{\small
\begin{align}
\mathbf{B}_{\mathrm{rh},n}^i &= \left[ \mathbf{G}_{\mathrm{rh}}  \left( \mathbf{A}(\boldsymbol{\varphi}^i_n) \diamond \mathbf{A}(\boldsymbol{\theta}^i_n) \right) \right] \odot \mathbf{A}_{t}(\boldsymbol{\nu}^{i}_n) \in \mathbb{C}^{ M\times L_n^i} \label{eq:brh},  \\ 
\mathbf{B}_{\mathrm{rv},n}^{i} &= \left[ \mathbf{G}_{\mathrm{rv}} \left( \mathbf{A}(\boldsymbol{\varphi}^{i}_n) \diamond \mathbf{A}(\boldsymbol{\theta}^{i}_n) \right) \right] \odot \mathbf{A}_{t}(\boldsymbol{\nu}^{i}_n) \in \mathbb{C}^{ M\times L_n^i} \label{eq:brv},\\
\boldsymbol{\Gamma}_{\mathrm{h},n}^i& = \mathrm{diag}({\gamma}_{\mathrm{h},n}^i(l)), \, \, \,\, l \in [1, \, L_n^i] \, ,\\
\boldsymbol{\Gamma}_{\mathrm{v},n}^i &= \mathrm{diag}({\gamma}_{\mathrm{v},n}^i(l)) \, ,\\
\mathbf{B}_{\mathrm{f},n}^i &= \mathbf{G}_{\mathrm{f}}^i \mathbf{A}(-\boldsymbol{\tau}^i_n) \in \mathbb{C}^{N_{\mathrm{f}}\times L_n^i}. \label{eq:bf}
\end{align}
%}% 
Here, $\mathrm{h}$ and $\mathrm{v}$ represent horizontal and vertical polarization, respectively. The \glspl{eadf} $\mathbf{G}_{\mathrm{rh}}, \mathbf{G}_{\mathrm{rv}}\in \mathbb{C}^{M\times N_{\mathrm{a}} N_{\mathrm{e}}}$ and the system frequency response $\mathbf{G}_{\mathrm{f}}^i$ are available from the system and antenna array calibration \cite{rimax_richter}. $N_{\mathrm{a}}$ and $N_{\mathrm{e}}$ are the dimensions of the azimuth \gls{aoa} and elevation \gls{aoa} of the \gls{eadf}. The phase shift matrix $\mathbf{A}(\boldsymbol{\mu}^i_n) \in \mathbb{C}$ is given by
\begin{equation}
    \mathbf{A}(\boldsymbol{\mu}^i_n)\rmv \rmv = \rmv \rmv \rmv 
    \left[
    \begin{array} {@{}c@{\;}c@{\;}c@{}}
  e^{-j\lfloor{\frac{N'}{2}}\rfloor\mu_{1,1}}  & \ldots & e^{-j\lfloor{\frac{N'}{2,n}}\rfloor\mu_{L_n^i,n}} \\
  \vdots &\ddots & \vdots \\
  e^{j(\lceil{\frac{N'}{2}}\rceil-1)\mu_{1,n}} & \ldots & e^{j(\lceil{\frac{N'}{2,n}}\rceil-1)\mu_{L_n^i,n}}
    \end{array}
    \right]
    \rmv \rmv \in \rmv \rmv \rmv\mathbb{C}^{N'\times L_n^i}.
\end{equation}
Here, $\bm{{\mu}}^i_n=[{\mu}^i_{1,n},{\mu}^i_{2,n}, \ldots, {\mu}^i_{L_{n}^i,n}]$ is a structural parameter vector that represents \gls{sp} delay $\boldsymbol{\tau}^{i}_n=[{\tau}^{i}_{1,n},{\tau}^{i}_{2,n},\ldots, {\tau}^{i}_{L_{n}^i,n}]$, azimuth \gls{aoa} $\boldsymbol{\varphi}^{i}_n=[{\varphi}^{i}_{1,n},{\varphi}^{i}_{2,n},\ldots, {\varphi}^{i}_{L_{n}^i},n]$, or elevation \gls{aoa} $\boldsymbol{\theta}^{i}_n=[{\theta}^{i}_{1,n},{\theta}^{i}_{2,n},\ldots, {\theta}^{i}_{L_{n}^i,n}]$. They are normalized to $[ 0, 2\pi )$, $[ -\pi, \pi )$ and $[ 0, \pi]$ respectively \cite{rimax_richter}, $N'$ corresponds to the dimensions $N_{\mathrm{f}}$, $N_{\mathrm{a}}$, and $N_{\mathrm{e}}$, respectively. 

The movement between the \gls{bs} and the mobile agent induces a Doppler shift that generates phase rotation across the sequentially-switched receiving antennas \cite{rui_hrpe_2017}. The phase rotation can be expressed as 
\begin{equation}
    \mathbf{A}_{t}(\boldsymbol{\nu}^{i}_n)\rmv\rmv =\rmv\rmv \begin{bmatrix}
  e^{j2\pi\frac{t_1}{\Delta T}\nu_{1,n}^i}  & \ldots & e^{j2\pi\frac{t_1}{\Delta T}\nu_{L_n^i,n}^i} \\
  \vdots & \ddots & \vdots \\
  e^{j2\pi\frac{t_M}{\Delta T}\nu_{1,n}^i} & \ldots & e^{j2\pi\frac{t_M}{\Delta T}\nu_{L_n^i,n}^i}
\end{bmatrix} \rmv\rmv \in \rmv \mathbb{C}^{M\times L_n^i}.
\end{equation}
Here, $\boldsymbol{\nu}^{i}_n=[{\nu}^{i}_{1,n},{\nu}^{i}_{2,n},\ldots, {\nu}^{i}_{L_{n}^i,n}]$, and $\Delta T$ denotes the total switching time of the antenna array. 

The \gls{sp} weights ${\gamma}_{\mathrm{h},n}^i(l)$ and ${\gamma}_{\mathrm{v},n}^i(l) $ are defined as
\begin{align}
{\gamma}_{\mathrm{h},n}^i(l)&=b_{\mathrm{th},n}^i  \bar{{\gamma}}_{\mathrm{hh},n}^i(l) + b_{\mathrm{tv},n}^i  \bar{{\gamma}}_{\mathrm{vh},n}^i(l) \, ,\\
{\gamma}_{\mathrm{v},n}^i(l)&=b_{\mathrm{th},n}^i  \bar{{\gamma}}_{\mathrm{hv},n}^i(l) + b_{\mathrm{tv},n}^i  \bar{{\gamma}}_{\mathrm{vv},n}^i(l). 
\end{align}
Here, $b_{\mathrm{th},n}^i$ and $b_{\mathrm{tv},n}^i$ are the horizontal and vertical antenna polarization of the $i$th \gls{bs}, and $\bar{{\gamma}}_{\mathrm{hh},n}^i(l)$, $\bar{{\gamma}}_{\mathrm{vh},n}^i(l)$, $\bar{{\gamma}}_{\mathrm{hv},n}^i(l)$, and $\bar{{\gamma}}_{\mathrm{vv},n}^i(l)$ represent different polarization combinations between the transmitter and the receiver for the $l$th \gls{sp}, e.g., $\mathrm{hh}$ denotes the horizontal-to-horizontal combination. 

To reflect the geometric characteristics of the wireless channel, the received signal is transformed into the delay-angle domain representation as follows:
\begin{equation} \label{eq:2d_z}
\mathbf{F}^i(n) = \|{\mathbf{B}}^{\mathrm{H}} \mathbf{Y}^i(n)\mathbf{B}^{*}_{f^{\prime}}\|^2 \in \mathbb{R}^{N_{\mathrm{a}} N_{\mathrm{e}} \times N_{\mathrm{d}}}.
\end{equation} 
The odd rows of $\mathbf{B}$ come from those of $\mathbf{B}^{\prime}_{\mathrm{rv}}$, and the even rows come from those of $\mathbf{B}^{\prime}_{\mathrm{rh}}$. Here, $\mathbf{B}^{\prime}_{\mathrm{rh}}$ and $\mathbf{B}^{\prime}_{\mathrm{rv}}$ are similar to \cref{eq:brh,eq:brv} but contain all the azimuth grids $\boldsymbol{\varphi}^{\prime}=[{-\pi},{-\pi(N_{\mathrm{a}}-2)}/N_{\mathrm{a}},\ldots, \pi(N_{\mathrm{a}}-2)/N_{\mathrm{a}}]$ and elevation grids $\boldsymbol{\theta}^{\prime}=[0,{\pi}/N_{\mathrm{e}},\ldots, \pi(N_{\mathrm{e}}-1)/N_{\mathrm{e}}]$. The delay matrix $\mathbf{B}_{{\mathrm{f}}^{\prime}}$ is similar to \cref{eq:bf}, but with delay grids $\boldsymbol{\tau}^{\prime}=[0,{2\pi}/N_{\mathrm{f}},\ldots, 2\pi(N_{\mathrm{d}}-1)/N_{\mathrm{f}}]$  and $N_{\mathrm{d}}$ representing the largest potential \gls{mpc} delay value. 

With this processing, the delay-angle domain signal is obtained. At the same time, the data size is increased from $MN_{\mathrm{f}}$ to $N_{\mathrm{a}}N_{\mathrm{e}}N_{\mathrm{d}}$, which is beneficial for machine learning by providing a large input space for pattern discovery \cite{studer2018channel}, at the cost of higher computational complexity.

%% file: 3_cellavit_structure.tex
\subsection{Single \gls{bs} \gls{vit} Structure}
$\mathbf{F}^i$ in \cref{eq:2d_z} can be rearranged into a 3D matrix as  ${\mathbf{F}^{i}}\in \mathbb{R}^{N_{\mathrm{a}} \times N_{\mathrm{e}} \times N_{\mathrm{d}}}$, which consists of delay, azimuth, and elevation dimensions. The index $n$ is dropped for notational convenience. Since the values of the elements in the matrix are large and vary greatly, they are not suitable for direct \gls{dnn} processing. Therefore, normalization is performed for each snapshot to obtain $\overline{\mathbf{F}}^{i} = p_i\mathbf{F}^{i}$. Here $p_i = \sqrt{N_{\mathrm{a}} N_{\mathrm{e}} N_{\mathrm{d}}}/{\|{\mathbf{F}}^{i}\|_{\mathrm{F}}}$ is the normalization factor for the $i$-th \gls{bs}.

\begin{figure}
	\centering    
    {\includegraphics[width=0.9\columnwidth]{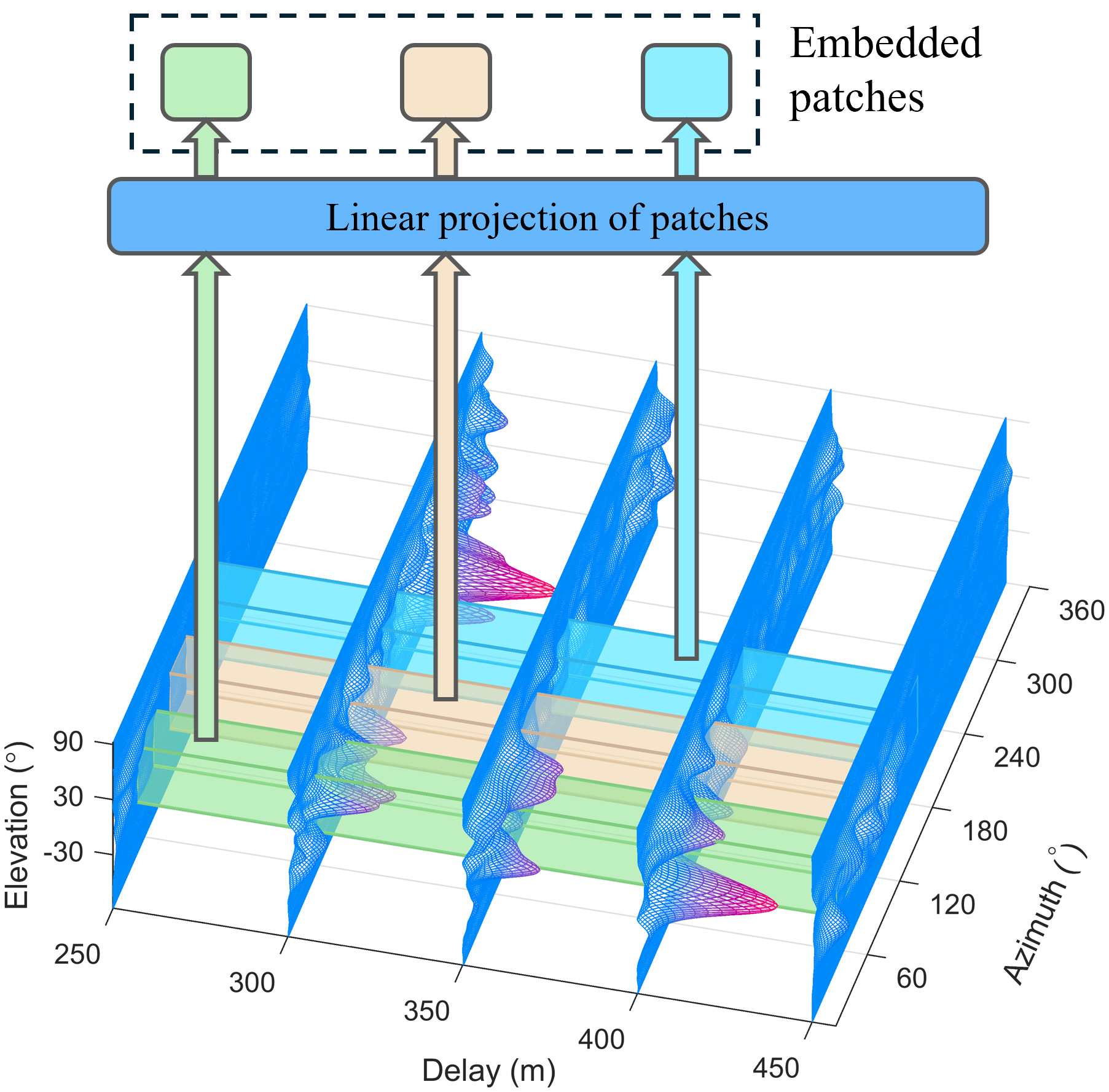}}
    \caption{The power distribution of $\overline{\mathbf{F}}^i$ at different delay, azimuth, and elevation grids, which is extracted from a representative real-world measurement snapshot. The patch embedding process of \gls{vit} is also illustrated.}
    \label{fig:patch_embed}
    \vspace{-5mm}
 \end{figure}

\cref{fig:patch_embed} illustrates the power distribution of $\overline{\mathbf{F}}^i$ at delay, azimuth, and elevation grids, which is extracted from a representative real-world measurement snapshot. It clearly shows that the received signal contains several \glspl{mpc} with distinct delay-angle properties. These geometric characteristics motivate and justify the application of image-based \glspl{dnn}, such as \glspl{vit} and \glspl{cvit}, as they allow the model to leverage the delay-angle correlation for high-precision positioning.   

Similar to the \gls{vit} architecture proposed in \cite{dosovitskiy2021vit}, the matrix $\overline{\mathbf{F}}^i$ can be segmented into several patches, and each patch $\mathbf{P}$ contains $P_{\mathrm{a}}, P_{\mathrm{e}}$, and $P_{\mathrm{d}}$ elements in the azimuth, elevation, and delay dimensions, respectively. The total number of patches is $K={N_{\mathrm{a}}N_{\mathrm{e}}N_{\mathrm{d}}}/(P_{\mathrm{a}} P_{\mathrm{e}}P_{\mathrm{d}})$. The patches are linearly embedded into $d_{\mathrm{model}}$ dimensions using a trainable projection matrix $\mathbf{E}\in \mathbb{R}^{(P_{\mathrm{a}}P_{\mathrm{e}}P_{\mathrm{d}}) \times d_{\mathrm{model}}}$ as shown in \cref{fig:patch_embed}. The output of the linear projection is termed a patch embedding or a token, given as
 \begin{equation}
    \mathbf{Z}_0 = \left[ \mathbf{P}_1 \mathbf{E};\, \mathbf{P}_2 \mathbf{E}; \, \cdots;\, \mathbf{P}_K \mathbf{E}\right] + \mathbf{E}_{\mathrm{pos}} 
\end{equation}
where $\mathbf{E}_{\mathrm{pos}}\in \mathbb{R}^{K\times d_{\mathrm{model}}}$ is the positional embedding applied to retain the positional information of each patch \cite{Vaswani2017attention,guoda2024attention}. 
 
\begin{figure}
	\centering	%\input{images/new_img/transformer_structure.tex}
    \includegraphics[width=0.7\columnwidth]{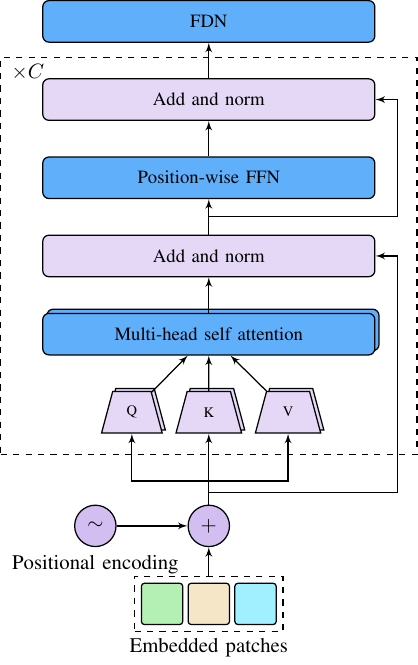}
    \caption{The \gls{vit} architecture for a single \gls{bs}.}
    \label{fig:cellavit}
    \vspace{-7mm}
\end{figure}

The sequence of embedded patches serves as the input to a Transformer encoder with $C$ layers of alternating \gls{msa} and position-wise \gls{ffn}. \Gls{ln} is applied after each block. The output is fed into a \gls{fdn} to generate the final estimates $\hat{\bm{p}}$, including the mobile agent's position and orientation. The process is shown as:
\begin{align}
    \mathbf{Z}^{\prime}_c &= \mathrm{LN}(\mathrm{MSA}(\mathbf{Z}_{c-1}) + \mathbf{Z}_{c-1}) \, \, \, \, c\in [1,C] \, ,\\
    \mathbf{Z}_{c} &= \mathrm{LN}(\mathrm{MLP}(\mathbf{Z}^{\prime}_{c}) + \mathbf{Z}^{\prime}_{c}) \, \, \, \, c\in [1,C]  \, ,\\
    \hat{\bm{p}} &= \mathrm{FCNN}(\mathbf{Z}_C).
\end{align}

The encoder architecture of \gls{vit} for a single \gls{bs} is shown in \cref{fig:cellavit}. The detailed implementation of each block is described as follows. First, \gls{msa} introduces self-attention among the patches and is implemented as:

\begin{list}{}{
    \setlength{\leftmargin}{0pt}
    \setlength{\itemindent}{0pt}
    \setlength{\labelwidth}{0pt}
    \setlength{\labelsep}{0pt}
}
    
\item \textbf{QKV linear projections}: the input $\mathbf{Z}$ is linearly transformed into queries $\mathbf{Q}$, keys $\mathbf{K}$, and values $\mathbf{V}$  with learnable weights $\mathbf{W}_{\mathrm{Q}}, \mathbf{W}_{\mathrm{K}}, \mathbf{W}_{\mathrm{V}} \in \mathbb{R}^{d_{\mathrm{model}} \times d_{\mathrm{model}}}$ and $\mathbf{Q},\mathbf{K}, \mathbf{V}\in \mathbb{R}^{K\times d_{\mathrm{model}}}$,
    \begin{equation}
        \mathbf{Q} = \mathbf{ZW}_{\mathrm{Q}}, \quad \mathbf{K} = \mathbf{ZW}_{\mathrm{K}}, \quad \mathbf{V} = \mathbf{ZW}_{\mathrm{V}}.
    \end{equation}
    
\item \textbf{Scaled dot-product attention (SDPA)}: the attention scores between queries and keys are computed, normalized by a softmax, and finally weighted and summed as: 
    \begin{equation}
        \mathrm{Attention}(\mathbf{Q},\mathbf{K},\mathbf{V}) = \mathrm{softmax}\left({\mathbf{Q}\mathbf{K}^\mathrm{T}}/{\sqrt{d_k}}\right)\mathbf{V}.
    \end{equation}
    Here, ${\mathbf{Q}\mathbf{K}^{\mathrm{T}}}/{\sqrt{d_k}}\in \mathbb{R}^{K\times K}$ represents the similarity matrix, and the softmax produces non-negative weights whose rows sum to $1$.  
\item \textbf{Multi-head self attention}: In MSA, the input data are split into $\mathrm{H}$ heads, and SDPA is applied to each head in parallel. This allows the model to jointly attend to information from different representation subspaces at different positions:
    
    \begin{equation}
        \mathrm{head}^{h} = \mathrm{Attention}(\mathbf{QW}_{\mathrm{Q}}^h, \mathbf{KW}_{\mathrm{K}}^h, \mathbf{VW}_{\mathrm{V}}^h)     \, .  
    \end{equation}
    Here $\mathbf{W}_{\mathrm{Q}}^h, \mathbf{W}_{\mathrm{K}}^h, \mathbf{W}_{\mathrm{V}}^h\in \mathbf{R}^{d_{\mathrm{model}}\times d_{\mathrm{model}}/H}$. 
    The concatenated heads are passed through a linear layer $\mathbf{W}^{\mathrm{O}}$ to project the fused information back into the model's dimension as:
\begin{equation}
    \mathrm{MSA}(\mathbf{Q}, \mathbf{K}, \mathbf{V}) = \mathrm{Concat}(\mathrm{head}^1, \dots, \mathrm{head}^H)\mathbf{W}^{\mathrm{O}}. 
\end{equation}
\end{list}

\gls{ln} is applied to each token vector $\bm{z}_k$ independently after \gls{msa} to stabilize and speed up training.

In the second step, the \Gls{mlp} serves as a Position-Wise \gls{ffn} and is presented as: 
\begin{equation}
    \mathrm{MLP}(\mathbf{Z}^{\prime}) = \max(0, \mathbf{Z}^{\prime}\mathbf{W}_1 + \mathbf{B}_1)\mathbf{W}_2 + \mathbf{B}_2 \, .
\end{equation}
Here $\mathbf{W}_1\in \mathbb{R}^{d_{\mathrm{model}}\times d_{\mathrm{hidden}}}, \mathbf{W}_2\in \mathbf{R}^{ d_{\mathrm{hidden}}\times d_{\mathrm{model}}}$ are learnable weights, and $\mathbf{B}_1\in \mathbf{R}^{K\times d_{\mathrm{hidden}}}, \mathbf{B}_2\in \mathbf{R}^{K\times d_{\mathrm{model}}} $ are learnable biases. 

In the third step, an \gls{fdn} is adopted to perform the mapping from the high-dimensional latent space to the physical coordinate manifold, and it is implemented by a multilayer \gls{fcnn} as:
\begin{align}    
\bm{v}_{0} &= \mathrm{vec}(\mathbf{Z}_C)  \, ,\\
\bm{v}_{c^{\prime}} &= \mathrm{ReLU} \left( \mathrm{LN} \left( \mathbf{W}_{c^{\prime}} \bm{v}_{c^{\prime}-1} + \bm{b}_{c^{\prime}} \right) \right) \, \,  c^{\prime} = [1, C^{\prime}]  \, ,\\
\hat{\bm{p}} &= \tanh(\mathbf{W}_{\mathrm{out}} \bm{v}_{C^{\prime}} + \bm{b}_{\mathrm{out}}).
\end{align}
The output $\hat{\bm{p}}$ contains both the position and yaw estimates. 

Finally, the cost function used to calculate the loss between the training output and the ground truth is implemented as:
\begin{equation} \label{eq:loss_function}
    J = \frac{1}{N_{\mathrm{tr}}}\sum_{n =1}^{N_{\mathrm{tr}}} (\alpha \| \overline{\bm{p}}_n - \hat{\bm{p}}_n\|^2  + \beta \| \overline{\bm{o}}_n-\hat{\bm{o}}_{n}\|^2). 
\end{equation}
Here $N_{\mathrm{tr}}$ is the total number of training snapshots, $\overline{\bm{p}}_n=[\frac{p_{\mathrm{x},n}}{p_{\mathrm{x},\max}}, \frac{p_{\mathrm{y},n}}{p_{\mathrm{y},\max}}]^{\mathrm{T}}$ is the normalized $\mathrm{x}, \mathrm{y}$ positions with $p_{\mathrm{x},\max}=\max(|p_{\mathrm{x},n}|)$ and $p_{\mathrm{y},\max}=\max(|p_{\mathrm{y},n}|), n\in[1, N_{\mathrm{tr}}]$, and $\overline{\bm{o}}_{n} = [\cos(\psi_n), \sin(\psi_n)]^{\mathrm{T}}$ is the normalized yaw of the mobile agent. The height, pitch and roll angles are treated as negligible in the absence of steep slopes and not estimated in the current setting. The cosine and sine representations avoid the yaw wrapping problem. Meanwhile, the normalized position loss has a magnitude comparable to that of the normalized yaw loss, making it easier and more flexible to select and balance the loss weights $\alpha$ and $\beta$. 

\subsection{Two \glspl{bs} \gls{vit} Structure}
\begin{figure}[!htbp]
	\centering    
    \includegraphics[width=0.65\columnwidth]{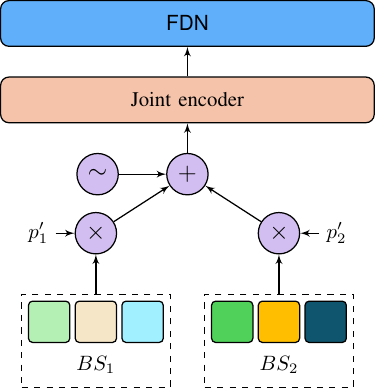}
    \caption{The joint \gls{vit} architecture for two \glspl{bs}.}
    \label{fig:joint_2bs}	
    \vspace{-7mm}
\end{figure}

For the two \glspl{bs} configuration, several combinations and processing strategies are possible. The signals can be processed either jointly or independently. In the joint processing method, the embeddings from two \glspl{bs} are passed to a unified, larger encoder for joint processing. The encoder output is fed into a larger \gls{fdn}. The system diagram is shown in \cref{fig:joint_2bs}. This approach enables the model to capture cross attention between two \glspl{bs} from the earliest stage. In this configuration, the total number of patches is $K={IN_{\mathrm{a}}N_{\mathrm{e}}N_{\mathrm{d}}}/(P_{\mathrm{a}} P_{\mathrm{e}}P_{\mathrm{d}})$, and the signals from two \glspl{bs} are normalized jointly before being sent to the encoder. The normalization is achieved as $\overline{\overline{\mathbf{F}}}^{i} =p^{\prime}_i \overline{\mathbf{F}}^{i} $, with
\begin{align}\label{eq:single_bs_norm}     
    p^{\prime}_i &= \sqrt{\frac{I}{\sum_{j=1}^I p_{j}^2}}\prod_{j=1,j\neq i}^I{p_{j}}.
\end{align}

\begin{figure}[!htbp]    
	\centering    
\includegraphics[width=0.65\columnwidth]{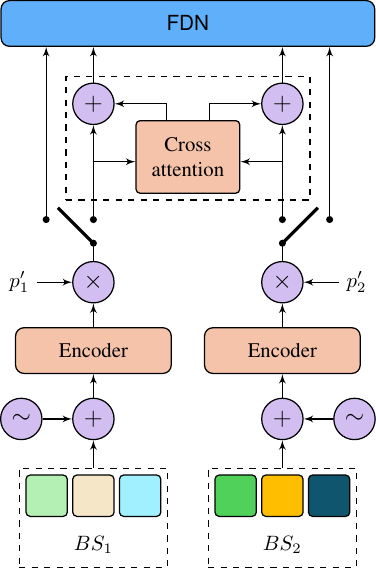}
    \caption{The independent \gls{vit} architecture for two \glspl{bs}.}
    \label{fig:indep_2bs}	
    \vspace{-8mm}
\end{figure}

Conversely, the independent processing method passes the embeddings from each \gls{bs} to a standard encoder and processes them independently. The encoder outputs are normalized afterward. The detailed diagram is shown in \cref{fig:indep_2bs}. This architecture can minimize mutual influence between the data streams of different \glspl{bs}, thereby improving the model's robustness. If the cross attention mechanism between two \glspl{bs} is disabled, the weighted outputs are directly concatenated and fed into a larger \gls{fdn}. Otherwise, they are sent to the cross attention node, whose outputs are added to the weighted outputs, and then concatenated and fed to the \gls{fdn}.

$\mathrm{Attention}(\mathbf{Q}_i,\mathbf{K}_{i^{\prime}},\mathbf{V}_{i^{\prime}})$ represents the cross attention from \gls{bs} $i$ to \gls{bs} $i^{\prime}$ with
\begin{equation}
        \mathbf{Q}_i = \mathbf{Z}_i\mathbf{W}_{\mathrm{Q}},  \quad \mathbf{K}_{i^{\prime}} = \mathbf{Z}_{i^{\prime}}\mathbf{W}_{\mathrm{K}}, \quad \mathbf{V}_{i^{\prime}} = \mathbf{Z}_{i^{\prime}}\mathbf{W}_{\mathrm{V}} \, .
    \end{equation}    
Similarly, the attention from \gls{bs} $i^{\prime}$ to \gls{bs} $i$ can be obtained. 

\subsection{\gls{cvit} Structure}
\begin{figure}[!htbp]
	\centering
    \includegraphics[width=0.65\columnwidth]{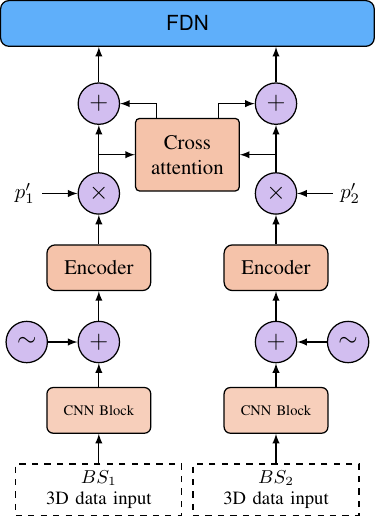}
    \caption{The proposed \gls{cvit} architecture for two \glspl{bs}.}
    \label{fig:convit}
    \vspace{-8mm}
\end{figure} 
In complex urban environments, the mobile agent may receive only \glspl{nlos} signals with low \gls{snr}, and the situation worsens when residual \gls{ici} degrades the signal quality. In such cases, the delay-angle representation is noisy and unsuitable for direct embedding. Preprocessing using a \gls{cnn} is desirable to capture local features and suppress noise and interference effectively. This provides a promising approach to generate high-quality input for the Transformer, which can capture the global representation of the environment at a later stage. A \gls{cvit} network is proposed, as shown in \cref{fig:convit}, to handle such challenging scenarios. The proposed network is similar to the \gls{vit} architecture, but the embedding stage is replaced and further enhanced by the \gls{cnn} block. The details of the \gls{cnn} block are introduced in \cref{subsec:network_structure}. The separated signal from each \gls{bs} is processed independently by one \gls{cnn} block, since a \gls{cnn} primarily captures local representations. The positional encoder is applied to the output of each \gls{cnn} block, and the results are sent to separate encoders. Cross attention is applied to the encoder outputs to capture the relationship between the two \glspl{bs}, and the outputs are sent to an \gls{fdn}. Due to the sparsity of the \gls{mpc} after the \gls{cnn} processing stage, a relatively shallow Transformer is sufficient to capture global attention.

%% file: 4_ekf.tex
Here, an \gls{ekf} \cite{GrovesPrinciples, thrun2005pr} is adopted to fuse high-frequency gyroscope and odometer data with the Transformer output to filter the position and orientation jointly. This approach is suitable for nonlinear motion, such as turning and acceleration. It focuses on trajectory smoothing and outlier rejection to mitigate the long tail outliers often present in \gls{dnn}-based predictions while providing real-time output.
\subsection{State Representation and Kinematic Model}
The system state at snapshot $n$ is defined by the vector $\tilde{\bm{x}}_n = [\tilde{p}_{\mathrm{x},n}, \tilde{p}_{\mathrm{y},n}, \tilde{\psi}_n]^{\mathrm{T}}$, representing the 2D Cartesian coordinates and the yaw angle. The control inputs at snapshot $n-1$ are $\bm{u}_{n-1} = [v_{\mathrm{odom},n-1}, \dot{\psi}_{\mathrm{gyro},n-1}]^{\mathrm{T}}$, where $v_{\mathrm{odom},n-1}$ is the longitudinal speed from the odometer and $\dot{\psi}_{\mathrm{gyro},n-1}$ is the angular velocity from the gyroscope. The nonlinear state transition function $f(\tilde{\bm{x}}_{n-1}, \bm{u}_{n-1})$ is given by:

\begin{equation}
\tilde{\bm{x}}_{n|n-1} = \begin{bmatrix} 
\tilde{p}_{\mathrm{x},n-1} + v_{\mathrm{odom},n-1} \cos(\tilde{\psi}_{n-1}) \Delta T \\
\tilde{p}_{\mathrm{y},n-1} + v_{\mathrm{odom},n-1} \sin(\tilde{\psi}_{n-1}) \Delta T \\
\tilde{\psi}_{n-1} + \dot{\psi}_{\mathrm{gyro},n-1} \Delta T
\end{bmatrix}.
\end{equation}

To propagate the state covariance $\mathbf{P}$, the transition is linearized using the Jacobian matrix $\mathbf{F}_{n-1}$:
\begin{equation}
\mathbf{F}_{n-1}
\rmv \rmv = \rmv \rmv \frac{\partial f}{\partial \tilde{\bm{x}}_{n-1}}
\rmv \rmv = \rmv \rmv 
\left[
\begin{array}{@{}c@{\;}c@{\;}c@{}}
1 & 0 & -v_{\mathrm{odom},n-1}\sin(\tilde{\psi}_{n-1})\Delta T \\
0 & 1 &  v_{\mathrm{odom},n-1}\cos(\tilde{\psi}_{n-1})\Delta T \\
0 & 0 & 1
\end{array}
\right]. 
\end{equation}
The predicted covariance is then updated as:
\begin{equation} 
\mathbf{P}_{n|n-1} = \mathbf{F}_{n-1} \mathbf{P}_{n-1|n-1} \mathbf{F}_{n-1}^{\mathrm{T}} + \mathbf{Q}, 
\end{equation}
where $\mathbf{Q}$ is the process noise covariance.

\subsection{Robust Measurement Update}
The unnormalized measurement from the Transformer is presented as:
\begin{equation}
\hat{\bm{z}}_n  = \big[\hat{p}_{\mathrm{x},n}p_{\mathrm{x},\max}, \, \hat{p}_{\mathrm{y},n}p_{\mathrm{y},\max},\, \mathrm{atan2}\big(\hat{o}_{n,1}, \hat{o}_{n,2}\big)\big]^{\mathrm{T}},
\end{equation}
and it is used to update the predicted state. The innovation $\bm{y}_n$ and its covariance $\mathbf{S}_n$ are computed as:
\begin{align}
\bm{y}_n &= \hat{\bm{z}}_n - \tilde{\bm{x}}_{n|n-1}  \, ,\\
\mathbf{S}_n &= \mathbf{P}_{n|n-1} + \mathbf{R}.
\end{align}
Here, $\mathbf{R}$ is the measurement noise covariance. Considering the non-Gaussian outliers in the Transformer outputs, the squared Mahalanobis distance is calculated as $d^2 = \bm{y}_n^{\mathrm{T}} \mathbf{S}_n^{-1} \bm{y}_n$. If $d^2$ exceeds the threshold $\gamma$, the measurement is rejected, and the filter defaults to dead reckoning. Otherwise, the filter computes the gain $\mathbf{K}_n$, which represents the optimal weighting between the kinematic prediction and the Transformer measurement:

\begin{equation}
\mathbf{K}_n = \mathbf{P}_{n|n-1}\mathbf{S}_n^{-1}.
\end{equation}

The state estimate $\tilde{\bm{x}}_{n|n}$ and error covariance $\mathbf{P}_{n|n}$ are then updated as:
\begin{align}
\tilde{\bm{x}}_{n|n} &= \tilde{\bm{x}}_{n|n-1} + \mathbf{K}_n \bm{y}_n  \, ,\\
\mathbf{P}_{n|n} &= (\mathbf{I} - \mathbf{K}_n) \mathbf{P}_{n|n-1}.
\end{align}

%% file: 5_measurements_setup.tex
\begin{figure}[!htbp] 
\centering
\scalebox{0.9}{\includegraphics[width=\linewidth]{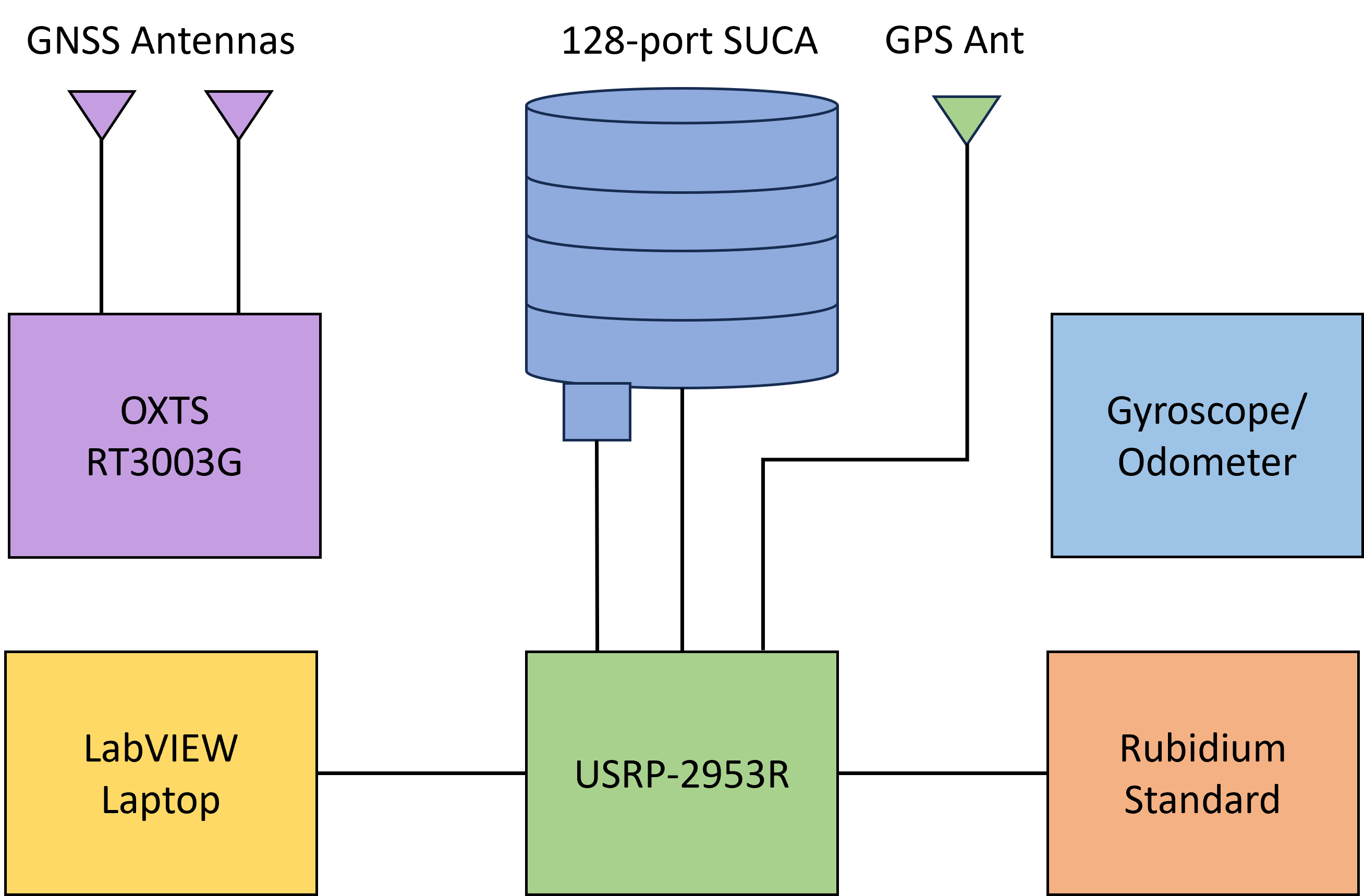}}
\caption{Block diagram of the measurement system and the ground truth system.}
\label{fig:measure_setup}
\vspace{-8mm}
\end{figure}

A measurement system was developed based on \gls{usrp}-2953R from National Instruments \cite{NIUSRP}. The block diagram of the measurement system is demonstrated in \Cref{fig:measure_setup}. It controlled the switching order of the $128$-port \gls{suca} and logged the time domain \gls{lte} signals from multiple \glspl{bs} to a laptop. The \gls{usrp} used a rubidium frequency standard \cite{Rubidium_Link}, which was disciplined by GPS beforehand, as a stable frequency reference to minimize absolute clock drift. The antenna array was mounted on the roof of a commercial vehicle, and the $128$ ports were switched in a predefined order controlled by the \gls{usrp}. The control signal had a $0.5$\,ms switching interval for each antenna port, and an additional $11$\,ms for \gls{agc} control, so the total time for one snapshot was $75$\,ms. The \gls{usrp}'s internal single antenna GPS receiver provided a one pulse per second (1 PPS) output to synchronize the \gls{usrp} with the other systems. It also logged its location as a comparison source. An OXTS RT3003G system \cite{oxts} generated the vehicle ground truth. It performed post-processed RTK with observation data from the SWEPOS network to provide the best possible pose estimates. The vehicle moved in the urban area of Lund, Sweden, at a relatively slow speed of around $\qty{1.0}{m/s}$ to avoid channel coherence limitations, which were imposed by the switch interval of the antenna array system. The total traveled time and distance were $32.5$\,minutes and $1750$\,meters. The vehicle also mounted a gyroscope and an odometer to acquire the longitudinal speed and the yaw of the antenna array. 

\begin{figure}[!hbtp]
	\centering	
	\hspace*{-2mm}
    \includegraphics[width=0.48\textwidth, height=0.42\textwidth]{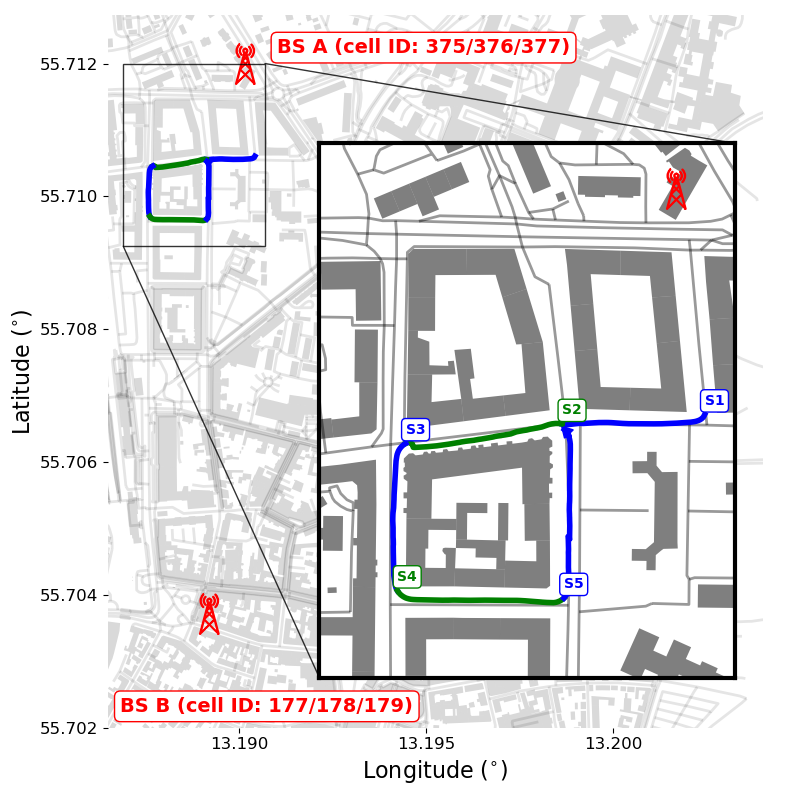}
    \caption{The vehicle's ground truth trajectory (S$1$ to S$5$) and the \glspl{bs} locations on the map of central Lund, Sweden. Adapted from \cite{junshi2026gmf}, licensed under CC BY 4.0.}
    \label{fig:ue_trajectory}
    \vspace{-6mm}
\end{figure}

The trajectory of the vehicle is shown in \Cref{fig:ue_trajectory}. It is divided into five segments (S1 to S5) with S2 to S5 constituting a closed loop. The vehicle starts from S1 and then moves counterclockwise along S2 to S5. Subsequently, it moves clockwise along S5 back to S2. The third lap is also clockwise, and the fourth lap is counterclockwise. The snapshot numbers for each lap are $5522$, $5626$, $5227$, and $4880$. There are two \glspl{bs} in the measurement field, both working at $2.66$\,GHz with $20$\,MHz bandwidth. \gls{bs} A (cell IDs $375/376/377$) is located north of the starting point of S1, at a distance of approximately $160$~m. It is almost completely blocked from the vehicle along the closed loop. \gls{bs} B (cell IDs $177/178/179$), is located to the south of the starting point of S5, at a distance of around $800$~m. Despite the long distance, it remains visible to the vehicle due to its height above the surrounding buildings. The colliding \glspl{crs} between two cells cause \gls{ici}, and necessitate \gls{icic} to separate the signals. 

\begin{figure}[!htbp]
    \centering
    \includegraphics[width=\columnwidth]{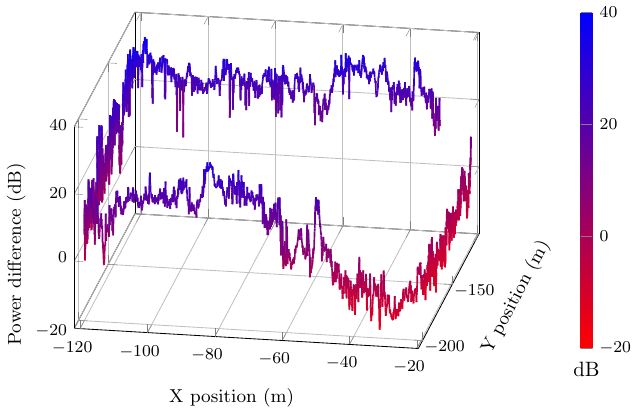}
    \caption{Power difference (in dB) between the colliding cells $376$ and $178$ after interference cancellation for lap $1$.}
    \label{fig:bs_power_ratio}
    \vspace{-9mm}
\end{figure}

The power difference in dB between the colliding cell $376$ and cell $178$ after interference cancellation for lap $1$ is shown in \cref{fig:bs_power_ratio}. Here, the figure only reflects the power difference between the baseband signals from these two cells at each snapshot, whereas the total power of the two cells at each snapshot is normalized to $1$. At segment S2, and most parts of S3 and S4, the signal from cell $376$ is dominant, but in most parts of S5, the signal from cell $178$ is dominant.

%% file: 6_measurement_results.tex
\subsection{Model Architecture and Complexity Comparison}\label{subsec:network_structure}
Three types of models are explored in this paper, including a standard \gls{resnet}, a \gls{vit} model, and the proposed \gls{cvit} model. Different combination strategies to handle signals from two \glspl{bs} using the \gls{vit} architecture are also explored. The details of the models are described in the following paragraphs.

\begin{figure}[!htbp]
	\centering
   \includegraphics[width=0.8\columnwidth]{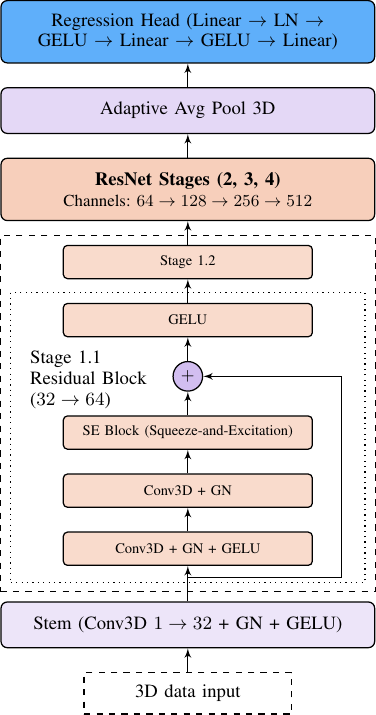}
    \caption{The reference 3D ResNet architecture.}
    \label{fig:resnet}
    \vspace{-8mm}
\end{figure}

A 3D \gls{resnet} is adopted as the first reference model. The detailed model architecture is shown in \cref{fig:resnet}. A \gls{resnet} backbone \cite{he2016deep} is employed and enhanced with Squeeze-and-Excitation (SE) blocks \cite{Hu2018SEN,hu2020SEN} for adaptive channel selection. To ensure stable training with small batch sizes for delay-angle domain data, group normalization \cite{wu2018groupnorm} is utilized throughout the architecture. All nonlinearities are implemented using the \gls{gelu} activation function \cite{hendrycks2016gelu} for improved gradient flow. The \gls{resnet} contains one stem layer, four \gls{resnet} stages, one adaptive average pooling layer, and one regression head. The stem layer produces 32 output channels, and the four \gls{resnet} stages, each containing two substages, gradually increase the output channels to $64,128,256$ and $512$ respectively, while simultaneously reducing the output dimensions. The adaptive average pooling layer converts the output dimension to $1$, and the regression head estimates the position and yaw values. More specifically, the regression head contains three linear layers to map the extracted features to the target output space. First, the input features ($d_{\mathrm{feat}}$, $512$ for single \gls{bs} and $1024$ for two \glspl{bs}) are projected into a latent space of dimension $d_{\mathrm{model}}$ via a linear layer followed by \gls{ln} and a \gls{gelu} activation function. A second linear layer reduces the dimensionality to $d_{\mathrm{model}}/4$. Finally, after a second \gls{gelu} activation, the last linear layer produces the position and yaw estimates.

For the two \glspl{bs} configuration, the input data are reshaped to effectively double the batch size. This approach ensures that the model architecture remains unchanged until the regression head, where the input dimensionality is doubled to accommodate the concatenated features. Consequently, the \gls{resnet} with a single \gls{bs} has nearly the same parameter count as the \gls{resnet} with two \glspl{bs}, despite the latter requiring twice the computational cost. Detailed architectural parameters, including kernel sizes, stride lengths, output channel numbers, and output dimensions, are provided in \cref{tab:resnet_para}.
\vspace{-5mm}
\newcolumntype{Y}{>{\centering\arraybackslash}X}
\newcolumntype{Z}{>{\centering\arraybackslash\hsize=0.9\hsize}X} % "S" for Small
\newcolumntype{L}{>{\centering\arraybackslash\hsize=1.4\hsize}X} % "L" for Large
\begin{table}[htbp]
\caption{Parameter list of the 3D \gls{resnet} }
\label{tab:resnet_para}
\centering
\begin{threeparttable}
\begin{tabularx}{\columnwidth}{|Z|Z|Z|Z|L|}
\hline
\textbf{Layer} & \makecell[c]{\textbf{Kernel} \\ \textbf{Size}} & \textbf{Stride} & \makecell[c]{\textbf{Output} \\ \textbf{Channels}} & \makecell[c]{\textbf{Output Dim.} \\ \textbf{($D, E, A$)}} \\
\hline
\rowcolor{blue!10!white} 
Input   & ---         & ---         & 1   & (100, 10, 180)\tnote{a} \\
Stem    & $(3, 2, 5)$ & $(2, 2, 2)$ & 32  & (50, 5, 90) \\
\rowcolor{blue!10!white} 
Stage 1.1 & $(3, 3, 3)$ & $(2, 1, 2)$ & 64  & (25, 5, 45) \\
Stage 1.2 & $(3, 3, 3)$ & $(1, 1, 1)$ & 64  & (25, 5, 45) \\
\rowcolor{blue!10!white}
Stage 2.1 & $(3, 3, 3)$ & $(2, 2, 2)$ & 128 & (13, 3, 23) \\
Stage 2.2 & $(3, 3, 3)$ & $(1, 1, 1)$ & 128 & (13, 3, 23) \\
\rowcolor{blue!10!white} 
Stage 3.1 & $(3, 3, 3)$ & $(2, 1, 2)$ & 256 & (6, 3, 12) \\
Stage 3.2 & $(3, 3, 3)$ & $(1, 1, 1)$ & 256 & (6, 3, 12) \\
\rowcolor{blue!10!white}
Stage 4.1 & $(3, 3, 3)$ & $(2, 1, 2)$ & 512 & (3, 3, 6) \\
Stage 4.2 & $(3, 3, 3)$ & $(1, 1, 1)$ & 512 & (3, 3, 6) \\
\rowcolor{blue!10!white} 
AvgPool & $Global$ & --- & 512 & (1, 1, 1) \\
\hline
\end{tabularx}
\begin{tablenotes}
\item[a] To reduce computational complexity, the elevation dimension of the input data is squeezed to 10 by taking the average of every 9 bins, which has been shown not to affect performance significantly \cite{russ2024wiometric}.
\end{tablenotes}
\end{threeparttable}
\vspace{-2mm}
\end{table}

A five-layer multi-head attention \gls{vit} model serves as the second benchmark. For the two \glspl{bs} configuration, two distinct processing strategies are evaluated. In the joint processing approach, as illustrated in \cref{fig:joint_2bs}, the signals from two \glspl{bs} are processed by a single encoder, resulting in twice the initial token number. Alternatively, in the independent processing scheme shown in \cref{fig:indep_2bs}, the signals are handled by separate encoders. In this case, the token count per encoder remains identical to the single \gls{bs} configuration, but the dimensionality of the input to the \gls{fdn} is doubled due to feature concatenation.

Furthermore, when cross attention is integrated into the independent architecture, the sum of the cross attention features and the encoder outputs is fed into the \gls{fdn}. The comprehensive architectural parameters for the \gls{vit} models are shown in \cref{tab:vit_para}.
\vspace{0mm}
\begin{table}[!htb]
    \centering
    \caption{Parameter list of the \gls{vit} model }
    \label{tab:vit_para}
    \renewcommand{\arraystretch}{1.3}     
    \begin{tabularx}{\columnwidth}{| >{\hsize=1.2\hsize}X | >{\centering\arraybackslash\hsize=0.8\hsize}X | >{\centering\arraybackslash\hsize=1.0\hsize}X |}    
    \hline
    \textbf{Hyperparameter} & \textbf{Symbol} & \textbf{Value}\\ 
    \hline
    \rowcolor{blue!10!white}      
    Model dimension             & $d_{\mathrm{model}}$ & 512 \\   
    FFN dimension               & $d_{\mathrm{hidden}}$   & 2048\\
    \rowcolor{blue!10!white}    
    Attention heads             & $\mathrm{H}$               & 8\\
    Head dimension              & $d_{\mathrm{model}}/\mathrm{H}$  & 64  \\
    \rowcolor{blue!10!white}      
    Number of layers            & $C$               & 5  \\ 
    Batch size                  & ---               & 50 \\
    \rowcolor{blue!10!white}
    Validation rate             & ---               & 0.15  \\      
    Dropout rate                & ---               & 0.1  \\
    \rowcolor{blue!10!white}
    Input data dimension        & $N_{\mathrm{d}}, N_{\mathrm{a}}, N_{\mathrm{e}}$   & $[100,10,180]$  \\
    Patch size                  & $P_{\mathrm{d}}, P_{\mathrm{a}}, P_{\mathrm{e}}$   & $[10,10,9]$     \\    
    \rowcolor{blue!10!white}
    Token number          & $n_{\mathrm{pos}}$  & 200, 400 (2 \glspl{bs}) \\    
    Loss weights & $\alpha, \beta$ & 1, 1        \\ 
    \hline
    \end{tabularx}
    \vspace{-2mm}
\end{table}

The proposed \gls{cvit} integrates \gls{cnn} and Transformer architectures. The \gls{cnn} backbone follows a design similar to that of \gls{resnet} except that it is streamlined to only two \gls{resnet} stages with modified kernel sizes and stride lengths to achieve a specific output channel number. Following the last convolutional stage, a projection layer maps the extracted features to the encoder inputs. After reshaping, the tokens maintain a dimensionality of $d_{\mathrm{model}}$ and the total token number is $13\times15=195$, which is comparable to the token number of \gls{vit} (200). The encoder part of the \gls{cvit} adopts the same parameters as the \gls{vit}, except that it is reduced to only two layers. The detailed parameters of the \gls{cnn} block are shown in \cref{tab:cvit_cnn_para}.
\vspace{0mm}
\begin{table}[htbp]
\caption{Parameter list of the \gls{cnn} block in the \gls{cvit}}
\label{tab:cvit_cnn_para}
\centering
\begin{tabularx}{\columnwidth}{|Z|Z|Z|Z|L|}
\hline
\textbf{Layer} & \makecell[c]{\textbf{Kernel} \\ \textbf{Size}} & \textbf{Stride} & \makecell[c]{\textbf{Output} \\ \textbf{Channels}} & \makecell[c]{\textbf{Output Dim.} \\ \textbf{($D, E, A$)}} \\
\hline
\rowcolor{blue!10!white} 
Input   & ---         & ---         & 1   & (100, 10, 180) \\
Stem    & $(3, 2, 5)$ & $(2, 2, 3)$ & 32  & (50, 5, 60) \\
\rowcolor{blue!10!white} 
Stage 1.1 & $(3, 3, 3)$ & $(2, 2, 2)$ & 64  & (25, 3, 30) \\
Stage 1.2 & $(3, 3, 3)$ & $(1, 1, 1)$ & 64  & (25, 3, 30) \\
\rowcolor{blue!10!white}
Stage 2.1 & $(3, 3, 3)$ & $(2, 3, 2)$ & 128 & (13, 1, 15) \\
Stage 2.2 & $(3, 3, 3)$ & $(1, 1, 1)$ & 128 & (13, 1, 15) \\
\rowcolor{blue!10!white} 
Projection & $(1, 1, 1)$ &$(1, 1, 1)$ & 512 & (13, 1, 15) \\
\hline
\end{tabularx}
\vspace{-3mm}
\end{table}

The comparative analysis of parameter counts and computational complexity across different models is summarized in \cref{tab:complexity_compare}. The results reveal a fundamental trade-off between the dense spatial feature extraction of \gls{resnet} and the sparse relational modeling of \gls{vit}. The \gls{resnet} models exhibit high computational intensity. For instance, the single-BS configuration requires $13.94$ \glspl{gflop}, indicating that 3D convolutions across the delay-angle domain ($100 \times 10 \times 180$) constitute the primary bottleneck. Conversely, the \gls{vit} architectures require more parameters but less computational overhead. For example, the single-BS \gls{vit} requires only $6.77$ \glspl{gflop} despite having $42.02$ million parameters. This efficiency is achieved through the \gls{vit} patching strategy, which compresses volumetric data into a tokenized embedding before applying attention mechanisms, but it is achieved at the expense of losing fine-grained local representations. 

Regarding multiple \glspl{bs} integration, the \gls{resnet} model maintains a nearly constant parameter count while doubling the computational cost. For \gls{vit}, the independent encoder approach essentially doubles the parameters and computation due to the parallel processing streams, with an additional overhead of $2.1$ million parameters and $0.51$ \glspl{gflop} introduced by the cross attention module. In contrast, the joint encoder configuration achieves comparable computational complexity ($14.37$ \glspl{gflop}) to that of the independent setup while maintaining a more efficient parameter profile ($68.24\,\mathrm{M}$). This suggests that the quadratic scaling of self-attention on a concatenated joint sequence is more efficient than the overhead of multi-stream processing coupled with cross-stream interaction. 

Notably, the proposed \gls{cvit}, which uses independent encoder with cross attention, significantly reduces both the parameter count and computational complexity compared to \gls{vit} with the same architecture. The parameter reduction is attributed to the \gls{cnn} block, which has fewer parameters, while the lower computational complexity is due to the reduced dimensionality of the output at each convolutional stage in the \gls{cnn} block. Finally, these results demonstrate that while \gls{resnet} is optimized for high-throughput spatial filtering, \gls{vit} offers a more computationally efficient alternative for global attention in multi-base station environments at the cost of increased parameter counts. The proposed \gls{cvit} achieves a favorable balance, offering lower computational complexity with fewer parameters.

\begin{table}
\renewcommand{\arraystretch}{1.3}  
\caption{Model complexity comparison}
\label{tab:complexity_compare}
\centering
\begin{tabularx}{\columnwidth}{| >{\hsize=1.2\hsize}X | >{\centering\arraybackslash\hsize=1.2\hsize}X | >{\centering\arraybackslash\hsize=0.6\hsize}X |}
\hline
\textbf{Model} & \textbf{Parameter count (M)} & \textbf{GFLOPs} \\
\hline
\rowcolor{blue!10!white}  
\gls{resnet} 1 \gls{bs} & 33.34 & 13.94 \\
\gls{resnet} 2 \glspl{bs} & 33.60 & 27.87 \\

\rowcolor{blue!10!white}
\gls{vit} 1 \gls{bs} & 42.02 & 6.77 \\
\gls{vit} 2 \glspl{bs} indep. encoder & 84.10 & 13.53 \\
 
 \rowcolor{blue!10!white} 
 \gls{vit} 2 \glspl{bs} indep. encoder with cross atten. & 86.20 & 14.04 \\
  \gls{vit} 2 \glspl{bs} joint encoder  & 68.24 & 14.37 \\ 
  \rowcolor{blue!10!white} 
  \gls{cvit} 2 \glspl{bs} indep. encoder with cross atten. & 69.92 & 10.66 \\ 
\hline
\end{tabularx}
\vspace{-3mm}
\end{table}

For all models, the data from the last three laps are selected as training and validation data, and the data from the first lap are selected as test data. Among the last three laps, $85\%$ of the snapshots are randomly selected as training data, and the rest are selected as validation data. The performance of the reference \gls{resnet} and \gls{vit} models is investigated using both a single \gls{bs} and two \glspl{bs}.

\subsection{Model Performance Comparison}
\begin{figure}[!htbp]
	\centering   
\includegraphics{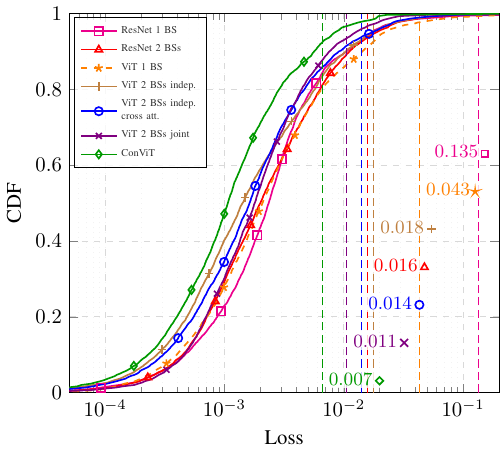}
    \caption{Loss \glspl{cdf} and \glspl{rmse} for different models.}
    \label{fig:loss_cdf}
    \vspace{-5mm}
\end{figure}

First, the \glspl{cdf} of the sample-wise loss defined in \cref{eq:loss_function} and the corresponding \gls{rmse} values using the test data are presented in \cref{fig:loss_cdf}. The \glspl{rmse} indicate that models using data from two \glspl{bs} achieve better performance than those using data from only one \gls{bs}. For the single \gls{bs} case, \gls{vit} outperforms \gls{resnet}. For the two \glspl{bs} cases, \gls{cvit} achieves the best performance, followed by \gls{vit} with joint encoder, \gls{vit} with independent encoder and cross attention, and \gls{resnet}. Finally, \gls{vit} with independent encoder achieves the worst performance. It indicates that \gls{cvit} suppresses local noise and effectively captures both local details and global attention, thus achieving the best performance. For \gls{vit} with a joint encoder, it captures cross attention between two \glspl{bs} from the beginning, thereby achieving better performance than \gls{vit} with independent encoder and cross attention, which captures cross attention only at a late stage. However, the latter is still slightly better than \gls{resnet} and outperforms \gls{vit} with independent encoder, which completely ignores cross attention. 

\begin{figure}[!htbp]
	\centering    
\includegraphics{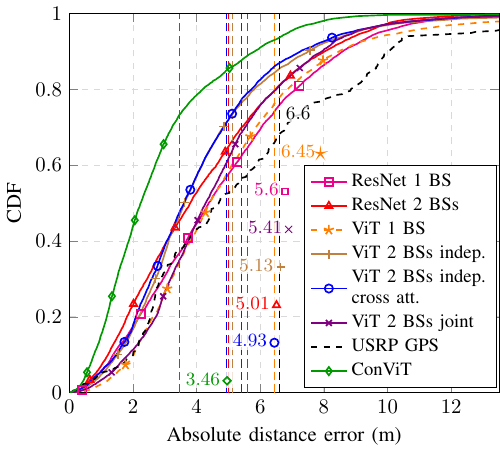}
    \caption{Absolute distance error \glspl{cdf} and \glspl{rmse} for different models.}
    \label{fig:dist_cdf}
    \vspace{-4mm}
\end{figure}

The \glspl{cdf} and \glspl{rmse} of the absolute distance error without filtering for different models are shown in \cref{fig:dist_cdf}. Additionally, the result from the USRP GPS Receiver (GPS L1) is also included. \gls{cvit} achieves the best performance with an \gls{rmse} of $3.46$ \,m. All distance results follow the same trend as the loss results, except that \gls{vit} with joint encoder performs worse than \gls{vit} with independent encoder. At the same time, \gls{vit} with a single \gls{bs} performs worse than \gls{resnet} with a single \gls{bs}. The performance degradation of the single antenna USRP GPS is attributed to the limited direct sky view in urban environments.

\begin{figure}[!htbp]
	\centering    
\includegraphics{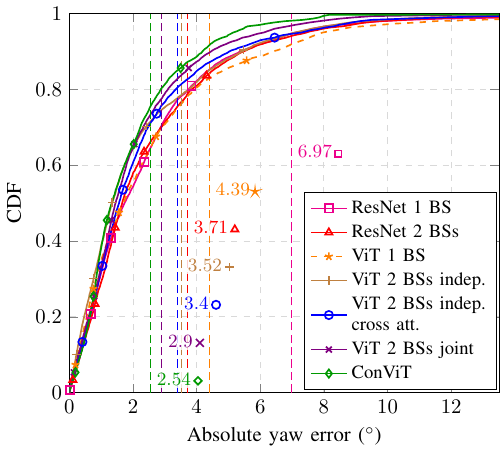}
    \caption{Absolute yaw error \glspl{cdf} and \glspl{rmse} for different models.}
    \label{fig:heading_cdf}
    \vspace{-9mm}
\end{figure}

The \glspl{cdf} and \glspl{rmse} of the absolute yaw error without filtering for different models are shown in \cref{fig:heading_cdf}. Since the GPS receiver is a point source and does not estimate yaw, its estimation is absent here. \gls{cvit} achieves the best performance with an \gls{rmse} of $2.54^{\circ}$. The yaw results follow the same trend as the loss results, except that \gls{resnet} with two \glspl{bs} performs worse than \gls{vit} with independent encoder. 

The \gls{rmse} values of loss, absolute distance error, and yaw error for all evaluated models are summarized in \cref{tab:rmse_list}. From \cref{fig:loss_cdf,fig:dist_cdf,fig:heading_cdf} and the table, it is observed that the models with two \glspl{bs} consistently outperform their single \gls{bs} counterparts since they can utilize more information. Furthermore, the proposed \gls{cvit} achieves the best performance across all metrics. By integrating \gls{cnn} stages before the Transformer encoder, \gls{cvit} effectively leverages the inductive bias of convolutional layers to extract fine-grained local features while utilizing the attention mechanisms of the encoder for global context. This hybrid approach ensures robust performance in both distance and yaw estimation. 

The \gls{vit} with independent encoder and cross attention demonstrates balanced performance across all metrics, although its individual metric \glspl{rmse} are worse than those of \gls{cvit}. It is also noticeable that \gls{vit} with joint encoder favors yaw estimation over distance estimation. This is likely due to the fact that yaw loss operates on a 1D manifold (the unit circle), presenting a more learnable target than 2D position coordinates. In the joint architecture, the self-attention mechanism appears to prioritize this lower-dimensional task to reduce the aggregate loss. The comparison between the joint and independent encoders highlights the role of architectural constraints. The independent encoder serves as a form of structural regularization by isolating \gls{bs} signals in the early layers. This may reduce the "gradient dominance" of the easier yaw estimation task and ensure that the model retains its representational capability and maintains balanced distance and yaw estimation performance.

\begin{table}[!htbp]
\vspace{-5mm}
\renewcommand{\arraystretch}{1.3}  
\caption{\glspl{rmse} of loss, distance, and yaw for different models}
\label{tab:rmse_list}
\centering
\begin{tabularx}{\columnwidth}{| >{\hsize=1.8\hsize}X | >{\centering\arraybackslash\hsize=0.6\hsize}X | >
{\centering\arraybackslash\hsize=1.0\hsize}X | >{\centering\arraybackslash\hsize=0.6\hsize}X |}
\hline
\textbf{Model} & \textbf{Loss} & \textbf{Distance (m)} & \textbf{Yaw ($^{\circ}$)} \\
\hline
\rowcolor{blue!10!white}  
\gls{resnet} 1 \gls{bs} & 0.135 & 5.6 & 6.97 \\
\gls{resnet} 2 \glspl{bs} & 0.016 & 5.01 & 3.71\\

\rowcolor{blue!10!white}
\gls{vit} 1 \gls{bs} & 0.043  & 6.45 & 4.39\\
\gls{vit} 2 \glspl{bs} indep. encoder & 0.018 & 5.13 & 3.52\\
 
 \rowcolor{blue!10!white} 
 \gls{vit} 2 \glspl{bs} indep. encoder with cross atten. & 0.014 & 4.93 & 3.4\\
  \gls{vit} 2 \glspl{bs} joint encoder  & 0.011 & 5.41 & 2.9\\ 
  \rowcolor{blue!10!white} 
  \gls{cvit} 2 \glspl{bs} indep. encoder with cross atten. & 0.007 & 3.46 & 2.54\\ 
  USRP GPS & - & 6.6 & -\\
\hline
\end{tabularx}
\vspace{-2mm}
\end{table}

\begin{figure}[!htbp]
    \centering
    \includegraphics[width=\columnwidth]{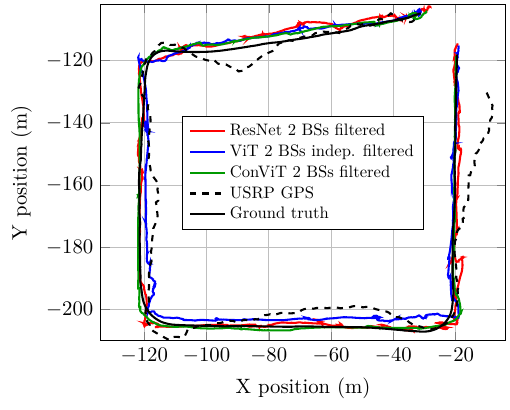}
    \caption{The filtered trajectories from \gls{resnet}, \gls{vit}, \gls{cvit}, the USRP GPS, and the ground truth.}
    \label{fig:model_trj}
    \vspace{-8mm}
\end{figure}

The \gls{ekf} filtered 2D trajectories of the \gls{resnet} with 2 \glspl{bs}, the \gls{vit} with independent encoder and cross attention, and the proposed \gls{cvit}, alongside the USRP GPS and the ground truth, are illustrated in \cref{fig:model_trj}. To enhance estimation stability, the \gls{ekf} incorporates speed and orientation information from an odometer and a gyroscope. It is observed that \gls{resnet} exhibits consistent error variance over different trajectory segments. In contrast, \gls{vit} outperforms \gls{resnet} in regions where a single \gls{bs} has dominant power by effectively leveraging global attention. However, its performance degrades in regions where two \glspl{bs} have comparable power (e.g., around $[-120, -190]$\,m and $[-40,-205]$\,m), likely due to increased residual \gls{ici} and the attention's sensitivity to less reliable feature representations in these areas. 

The proposed \gls{cvit} consistently outperforms both \gls{resnet} and \gls{vit} across the entire trajectory. By leveraging the \gls{cnn} block, \gls{cvit} successfully suppresses signal noise and extracts more reliable feature representations while maintaining global attention on these refined representations in the late stage. Finally, the USRP GPS demonstrates the worst performance due to the challenging urban environment with restricted sky visibility.
 
\begin{figure}[!htbp]
    \centering
    \includegraphics[width=\columnwidth]{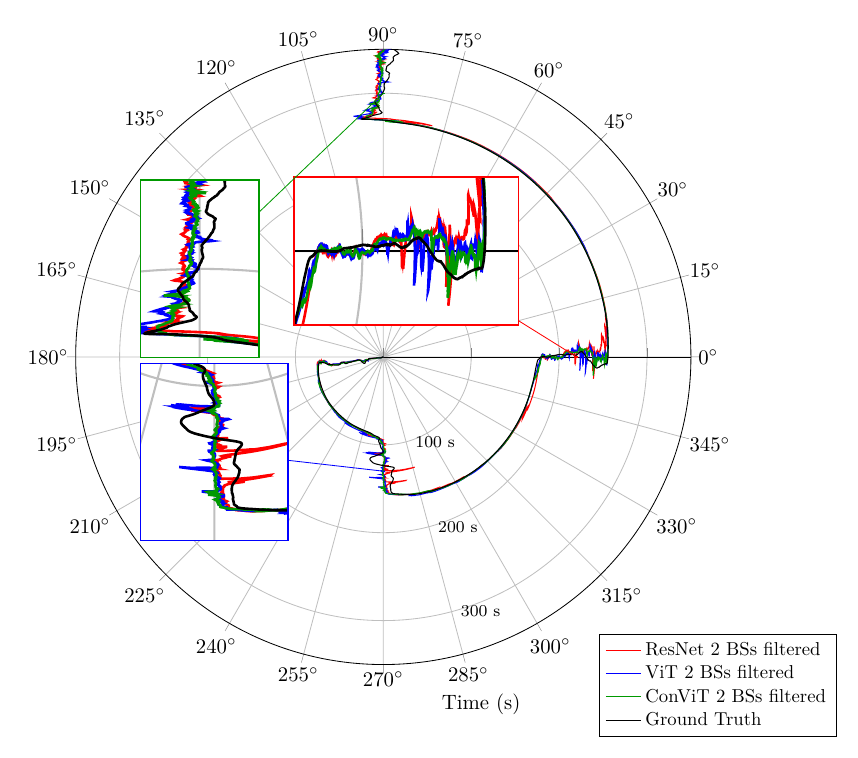}
    \caption{The filtered yaw estimates in polar coordinates from \gls{resnet}, \gls{vit}, \gls{cvit} and the ground truth.}
    \label{fig:model_yaw}
    \vspace{-5mm}
\end{figure}

The filtered yaw estimations for the aforementioned models and the ground truth are illustrated in \cref{fig:model_yaw} in polar coordinates. A primary observation is that all models demonstrate superior performance at turns compared with straight segments. While this is seemingly counterintuitive, given the fewer training snapshots available for turns, these areas provide higher yaw variability. The models effectively utilize this diversity to learn more robust directional features. Conversely, the nearly invariant yaw values along straight segments limit the model's ability to extract discriminative features, leading to degraded yaw estimation performance. 

The \gls{ekf} with sensor fusion plays a critical role in mitigating these instabilities. By fusing sensor measurements with the deep learning model outputs, the filter effectively reduces model uncertainty, especially at straight segments. By smoothing the long ``tail'' fluctuations produced by the models, the \gls{ekf} complements the models and provides a continuous and more stable yaw estimation. Consistent with previous observations, \gls{resnet} is outperformed by \gls{vit}, while \gls{cvit} achieves the highest yaw estimation accuracy. 

\begin{figure*}[!t]
    \centering
    % --- First Subfigure ---
    \subfloat{\label{subfig:snapshot1376}
    \includegraphics[width=0.9\textwidth, height=0.35\textwidth]{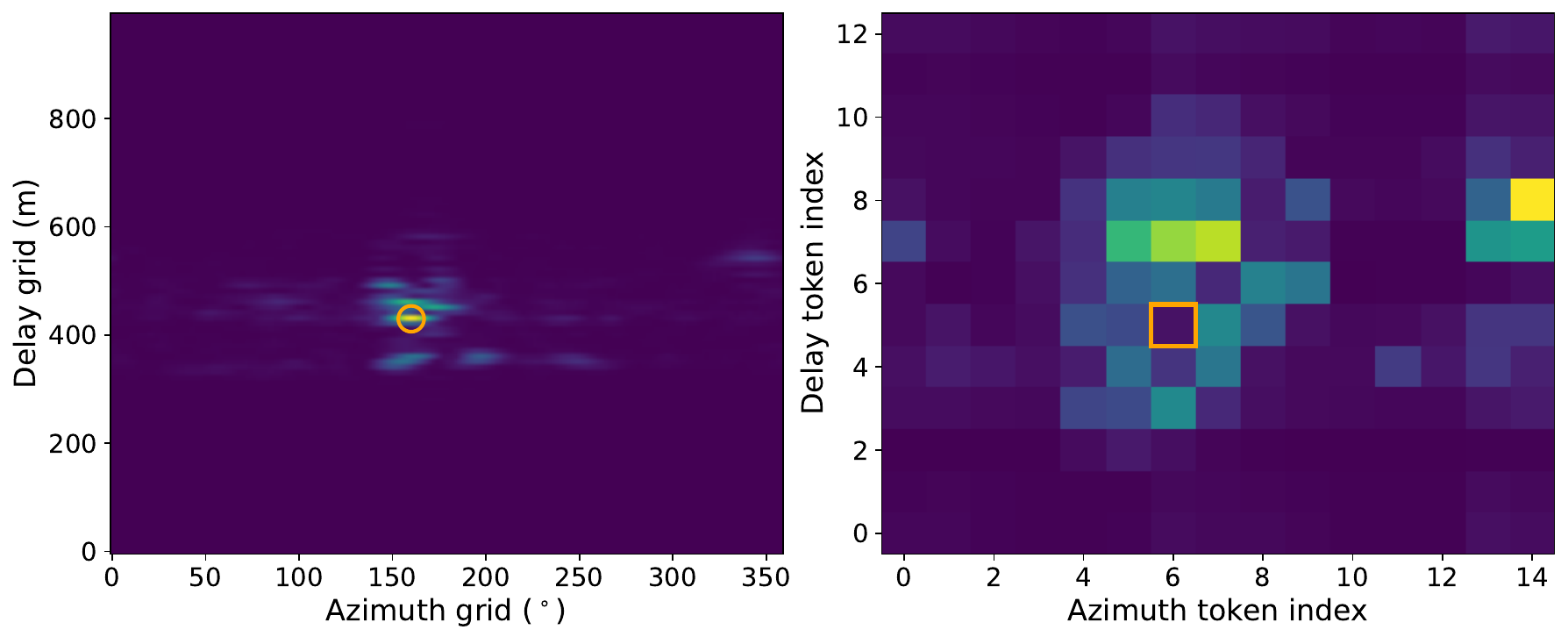}}
    \vspace{2mm} 
    % --- Second Subfigure ---
    \subfloat{\label{subfig:snapshot1888}
    \includegraphics[width=0.9\textwidth, height=0.35\textwidth]{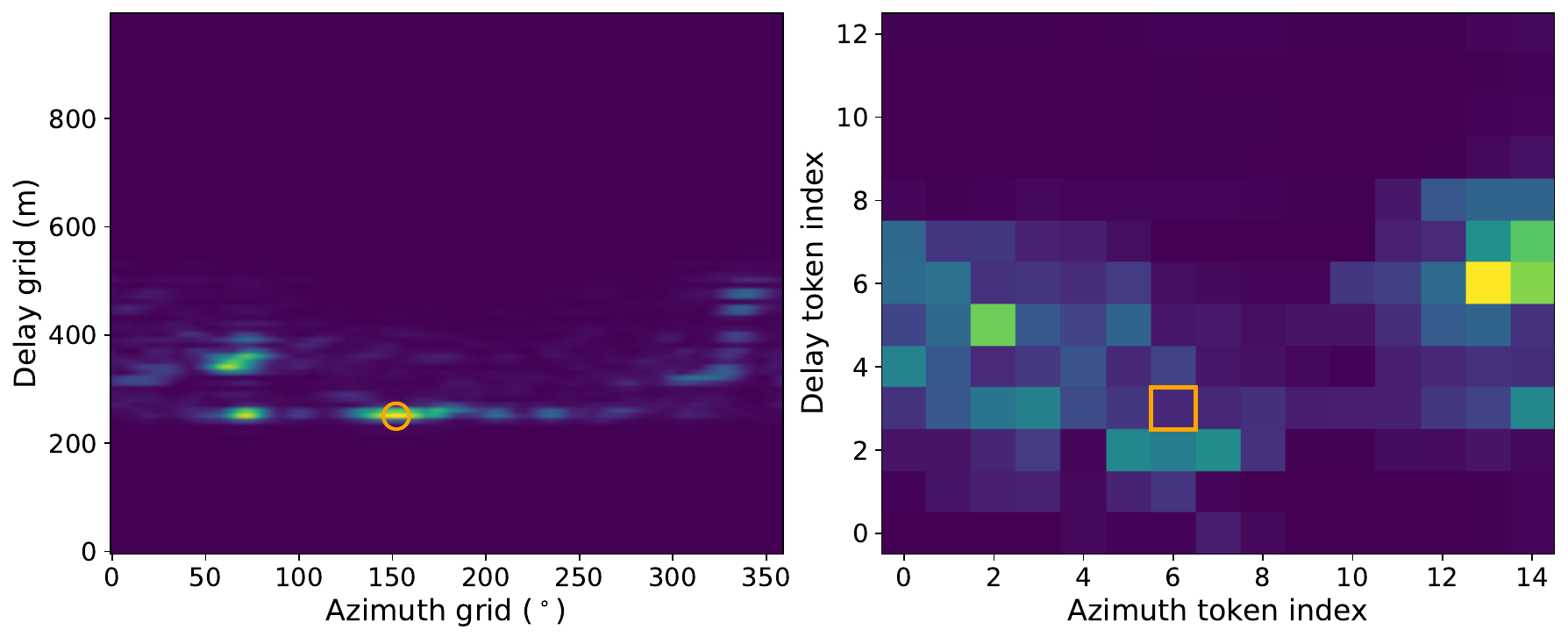}}
    
    \caption{Comparison of the input signal power distribution in the delay-angle domain from cell $376$ and the corresponding Transformer attention weights for interpretability analysis. (Left) Input signal power in the delay-angle domain for snapshots 1376 and 1888, with peak values indicated by orange circles. (Right) Corresponding last-layer attention weights from the $6$th head for the tokens with peak power (orange squares). Brighter colors denote higher power or attention weights.}
    \label{fig:tf_atten_w}
    \vspace{-7mm}
\end{figure*}

Finally, to evaluate the interpretability of the proposed model, the input signal power distribution of cell $376$ is compared with the corresponding Transformer's internal attention weights. As illustrated in \cref{fig:tf_atten_w}, there is a clear spatial correspondence between the original power distribution in the delay-angle domain and the resulting attention weight maps. Specifically, for snapshots 1376 and 1888, the identified peak power component (indicated by orange circles in the left subfigures) correlates with the focus of the attention mechanism. The attention weights for the corresponding tokens (indicated by orange squares in the right subfigures) to other tokens are nonuniformly distributed. Instead, the $6$th attention head of the last layer concentrates its weights on tokens aligned with the high-power regions. This topological consistency suggests that the Transformer mechanism inherently captures the physical structure of the wireless channels. By effectively attending to \glspl{mpc} with high power, the model provides evidence that its internal attention mechanism is physically related to the signal's energy distribution.

%% file: 7_conclusion.tex
This paper presents a hybrid \gls{ml} model, namely \gls{cvit}, for high-accuracy mobile agent positioning using cellular signals received in complex urban environments. The received signals are transformed into delay-angle domain representations, which reflect the geometric characteristics of the wireless channels and exhibit image-like structures, making them well suited to vision-based neural networks. 

By integrating the local feature refinement  capability of \gls{cnn} modules with the global dependency modeling capability of Transformers, the proposed \gls{cvit} architecture effectively addresses the challenges caused by noise and residual \gls{ici} in urban environments. It also fuses signals from two \glspl{bs} with power imbalance to ensure balanced and resilient performance in both position and yaw estimation. Furthermore, the integration of an \gls{ekf} demonstrates that sensor fusion can reliably mitigates the model uncertainty and enhances overall performance.  

The proposed \gls{cvit} model, together with \gls{resnet} and \gls{vit} models, is validated using field measurement data collected in Lund, Sweden. The experimental results demonstrate that \gls{cvit} achieves a distance \gls{rmse} of $3.46$ m and a yaw \gls{rmse} of $2.54^{\circ}$, significantly outperforming the reference models. Meanwhile, it maintains fewer parameters and lower computational complexity than the \gls{vit} model with the same architecture.

Finally, the interpretability analysis confirms that the Transformer's internal attention mechanisms are physically grounded, with some heads selectively attending to high-power \glspl{mpc} in the delay-angle domain. Even with limited training data, the \gls{cvit} model provides a promising solution for cellular signal-based positioning in challenging real-world environments. 

%% file: 8_acknowledgment.tex
The authors would like to thank Dr. Guoda Tian from Ericsson, Lund, for his help and insightful discussions during the course of the work.

%% file: IEEEabrv.bib
@STRING{IEEE_J_AES        = "{IEEE} Trans. Aerosp. Electron. Syst."}

@STRING{IEEE_J_ITS        = "{IEEE} Trans. Intell. Transp. Syst."}

@STRING{IEEE_J_VT         = "{IEEE} Trans. Veh. Technol."}

@STRING{IEEE_J_WCOM       = "{IEEE} Trans. Wireless Commun."}

@STRING{IEEE_OJ_COM       = "{IEEE} Open J. Commun. Soc."}

@STRING{IEEE_J_PAMI       = "{IEEE} Trans. Pattern Anal. Mach. Intell."}

@STRING{IEEE_J_PROC       = "Proc. {IEEE}"}

@STRING{IEEE_M_AES        = "{IEEE} Aerosp. Electron. Syst. Mag."}

@STRING{IEEE_M_COM        = "{IEEE} Commun. Mag."}

@STRING{IEEE_M_SP         = "{IEEE} Signal Process. Mag."}

@STRING{IEEE_M_VT         = "{IEEE} Veh. Technol. Mag."}

@STRING{IEEE_M_WC         = "{IEEE} Wireless Commun. Mag."}

@STRING{IEEE_O_CSTO        = "{IEEE} Commun. Surveys Tuts."}


%% file: liberry_short.bib
@InProceedings{LiLeiCaiTuf:ICC2024,
	author    = {Li, Xuhong and Cai, Xuesong and Leitinger, Erik and Tufvesson, Fredrik},
	booktitle = {Proc. {IEEE} {ICC} 2024},
	title     = {A Belief Propagation Algorithm for Multipath-based {SLAM} with Multiple Map Features: {A} {mmWave} {MIMO} Application},
	year      = {2024},
	month     = {Aug.},
	doi       = {10.1109/ICCWorkshops59551.2024.10615625},
	pages     = {269-275},	
}

@Article{LiFUSION2026,
      title={Probabilistic Occupancy Grid for Radio-Based {SLAM}}, 
      author={Xuhong Li and Erik Leitinger and Fredrik Tufvesson and Florian Meyer},
      year={2026},
      eprint={2603.03559},
      archivePrefix={arXiv},
      primaryClass={eess.SP},
      url={https://arxiv.org/abs/2603.03559}, 
}

@Article{venus2026FUSION,
      title ={{AI}-enhanced Direct {SLAM}: A Principled Approach to Unsupervised Learning in {B}ayesian Inference}, 
      author={Alexander Venus and Benjamin Deutschmann and Alexander Fuchs and Christian Knoll and Erik Leitinger},
      year  ={2026},
      eprint={2603.01071},
      journal={arXiv},
      primaryClass={eess.SP},
      url   ={https://arxiv.org/abs/2603.01071}, 
}

@Article{deutschmann2026TSP,
      title ={Soft-Coherent Direct Multipath {SLAM}}, 
      author={Benjamin J. B. Deutschmann and Klaus Witrisal and Erik Leitinger},
      year  ={2026},
      eprint={2604.19723},
      journal={arXiv},
      primaryClass={eess.SP},
      url   ={https://arxiv.org/abs/2604.19723}, 
}

@Article{LiTWC2026,
  author    ={Li, Xuhong and Deutschmann, Benjamin J. B. and Leitinger, Erik and Meyer, Florian},
  journal   = IEEE_J_WCOM,
  title     ={Adaptive Multipath-Based {SLAM} for Distributed {MIMO} Systems}, 
  year      ={2026},
  month     ={Apr.},
  volume    ={25},
  number    ={},
  pages     ={16931-16946},
  doi       ={10.1109/TWC.2026.3679142},
  }

@ARTICLE{harsh_6g_paper,  author={Tataria, Harsh and Shafi, Mansoor and Molisch, Andreas F. and Dohler, Mischa and Sjöland, Henrik and Tufvesson, Fredrik},
journal=IEEE_J_PROC,   title={{6G} wireless systems: vision, requirements, challenges, insights, and opportunities}, 
month={mar.},
year={2021},
volume={109},
number={7}, 
pages={1166-1199}, 
doi={10.1109/JPROC.2021.3061701}
}

@article{1g_to_5g_cellular_localization,
  title={Survey of cellular mobile radio localization methods: From {1G} to {5G}},
  author={del Peral-Rosado, Jos{\'e} A and Raulefs, Ronald and L{\'o}pez-Salcedo, Jos{\'e} A and Seco-Granados, Gonzalo},
  journal=IEEE_O_CSTO,
  volume={20},
  number={2},
  pages={1124--1148},
  month={Dec.},
  year={2017},
  publisher={IEEE},
  doi={10.1109/COMST.2017.2785181}
}

@ARTICLE{Saleh2026vehicularpositioning,
  author={Saleh, Sharief and Dwivedi, Satyam and Whiton, Russ and Hammarberg, Peter and Keskin, Musa Furkan and Equi, Julia and Chen, Hui and Munier, Florent and Eriksson, Olof and Gunnarsson, Fredrik and Tufvesson, Fredrik and Wymeersch, Henk},
  journal= IEEE_O_CSTO, 
  title={Vehicular wireless positioning: A survey}, 
  month={Mar.},
  year={2026},
  volume={28},
  number={},
pages={5792-5832},
  doi={10.1109/COMST.2026.3674599}
  }

@article{whiton2022cellular,
  title={Cellular localization for autonomous driving: A function pull approach to safety-critical wireless localization},
  author={Whiton, Russ},
  _old_journal={IEEE Vehicular Technology Magazine},
journal = IEEE_M_VT,
 month={Oct.},
  year={2022},
  volume={17},
  number={4},
  pages={28-37},
  publisher={IEEE},
    doi={10.1109/MVT.2022.3208392}
}

@ARTICLE{Ko2021v2xpos,
  author={Ko, Seung-Woo and Chae, Hyukjin and Han, Kaifeng and Lee, Seungmin and Seo, Dong-Wook and Huang, Kaibin},
  journal=IEEE_M_WC, 
  title={{V2X}-based vehicular positioning: Opportunities, challenges, and future Directions}, 
  month={mar.},
  year={2021},
  volume={28},
  number={2},
  pages={144-151},
  doi={10.1109/MWC.001.2000259}
  }

@article{dwivedi2021positioning,
  title={Positioning in {5G} Networks},
  author={Dwivedi, Satyam and Shreevastav, Ritesh and Munier, Florent and Nygren, Johannes and Siomina, Iana and Lyazidi, Yazid and Shrestha, Deep and Lindmark, Gustav and Ernstr{\"o}m, Per and Stare, Erik and others},  
  journal = IEEE_M_COM,
  %journal={IEEE Communications Magazine},
  volume={59},
  number={11},
  pages={38--44},
  month={Nov.},
  year={2021},
  publisher={IEEE},
doi={10.1109/MCOM.011.2100091}
}

@ARTICLE{kassas_magazine,
  author={Z. {Kassas} and J. {Khalife} and K. {Shamaei} and J. {Morales}},
  journal=IEEE_M_SP, 
  title={{I} hear, therefore {I} Know where {I} am: Compensating for {GNSS} limitations with cellular signals}, 
  month={Sept.},
  year={2017},
  volume={34},
  number={5},
  pages={111-124},  
  doi={10.1109/MSP.2017.2715363},
  ISSN={1558-0792},
  month={Sep.}
  }

@article{zhu2018gnss,
  title={{GNSS} position integrity in urban environments:
A Review of Literature},
  author={Zhu, Ni and Marais, Juliette and B{\'e}taille, David and Berbineau, Marion},
  journal=IEEE_J_ITS,
  _old_journal={IEEE Transactions on Intelligent Transportation Systems},
  volume={19},
  number={9},
  pages={2762--2778},
  month={September},
  year={2018},
  publisher={IEEE},
  doi={10.1109/TITS.2017.2766768}
}

@book{GrovesPrinciples,
author = {Groves, Paul},
year = {2013},
month = {Mar.},
pages = {},
address={Norwood, MA, USA},
title = {Principles of GNSS, Inertial, and Multisensor Integrated Navigation Systems, 2nd ed.},
publisher={Artech House}
}

@ARTICLE{Yang2022ltetriang,
  author={Yang, Chun and Arizabaleta-Diez, Markel and Weitkemper, Petra and Pany, Thomas},
  journal=IEEE_M_AES,
  _old_journal={IEEE Aerospace and Electronic Systems Magazine}, 
  title={An experimental analysis of cyclic and reference signals of {4G} {LTE} for {TOA} estimation and positioning in mobile fading environments}, 
  month={September},
  year={2022},
  volume={37},
  number={9},
  pages={16-41},
  doi={10.1109/MAES.2022.3186650}
  }

@ARTICLE{ChSLAMOrig,  
author={Gentner, Christian and Jost, Thomas and Wang, Wei and Zhang, Siwei and Dammann, Armin and Fiebig, Uwe-Carsten},
journal=IEEE_J_WCOM,
title={Multipath Assisted Positioning with Simultaneous Localization and Mapping},
month={Jun.},
year={2016},
volume={15},
number={9},
pages={6104-6117},
doi={10.1109/TWC.2016.2578336}}

@ARTICLE{Kim2024TSP,
  author    = {Kim, Hyowon and García-Fern{\'a}ndez, {\'A}ngel F. and Ge, Yu and Xia, Yuxuan and Svensson, Lennart and Wymeersch, Henk},
  journal   = {{IEEE} Trans. Signal Process.},
  title     = {Set-Type Belief Propagation With Applications to {P}oisson Multi-{B}ernoulli {SLAM}}, 
  year      = {2024},
  month     = {Apr.},
  volume    = {72},
  number    = {},
  pages     = {1989-2005},
  doi       = {10.1109/TSP.2024.3383543}}

@ARTICLE{pan2026aipossurvey,
  author={Pan, Guangjin and Gao, Yuan and Gao, Yilin and Yu, Wenjun and Zhong, Zhiyong and Yang, Xiaoyu and Guo, Xinyu and Xu, Shugong},
  journal=IEEE_O_CSTO, 
  title={{AI}-driven wireless positioning: Fundamentals, standards, state-of-the-art, and challenges}, 
  month={December},
  year={2026},
  volume={28},
  number={},
  pages={4394-4428},
  doi={10.1109/COMST.2025.3648577}
  }

@ARTICLE{Sze2017dnnpossurvey,
  author={Sze, Vivienne and Chen, Yu-Hsin and Yang, Tien-Ju and Emer, Joel S.},
  journal=IEEE_J_PROC,
  _full_journal={Proceedings of the IEEE}, 
  title={Efficient processing of deep neural networks: A tutorial and survey}, 
  month={December},
  year={2017},
  volume={105},
  number={12},
  pages={2295-2329},
 doi={10.1109/JPROC.2017.2761740}
  }

@article{krizhevsky2017imagenet,
  title={Imagenet classification with deep convolutional neural networks},
  author={Krizhevsky, Alex and Sutskever, Ilya and Hinton, Geoffrey E},
  journal={Commun. ACM},
  volume={60},
  number={6},
  pages={84--90},
  month={June},
  year={2017},
  publisher={AcM New York, NY, USA},
  doi = {10.1145/3065386},  
}

@inproceedings{joaovieira2017deep,
  title={Deep Convolutional Neural Networks for Massive {MIMO} Fingerprint-Based Positioning},
  author={Vieira, Joao and Leitinger, Erik and Sarajlic, Muris and Li, Xuhong and Tufvesson, Fredrik},
  booktitle={Proc. IEEE 28th Annu. Int. Symp. Pers, Indoor, and Mobile Radio Commun. (PIMRC)},
  _old_booktitle={2017 IEEE 28th Annual International Symposium on Personal, Indoor, and Mobile Radio Communications (PIMRC)},
  pages={},
  month={October},
  year={2017},
  organization={IEEE},
  address={Montreal, QC, Canada}
}

@ARTICLE{russ2024wiometric,
  author={Whiton, Russ and Chen, Junshi and Tufvesson, Fredrik},
  journal=IEEE_J_VT,
  _old_journal={IEEE Transactions on Vehicular Technology}, 
  title={Wiometrics: Comparative Performance of Artificial Neural Networks for Wireless Navigation}, 
  month={September},
  year={2024},
  volume={73},
  number={9},
  pages={13883-13897},  
  doi={10.1109/TVT.2024.3396286}}

@ARTICLE{sun2019fingerprint,
  author={X. {Sun} and C. {Wu} and X. {Gao} and G. Y. {Li}},
  journal=IEEE_J_VT,
  _old_journal={IEEE Transactions on Vehicular Technology}, 
  title={Fingerprint-Based Localization for Massive {MIMO-OFDM} System With Deep Convolutional Neural Networks}, 
  month={September},
  year={2019},
  volume={68},
  number={11},
  pages={10846-10857},
    doi={10.1109/TVT.2019.2939209}
  }

@inproceedings{Vaswani2017attention,
author = {Vaswani, Ashish and Shazeer, Noam and Parmar, Niki and Uszkoreit, Jakob and Jones, Llion and Gomez, Aidan N. and Kaiser, {\L}ukasz and Polosukhin, Illia},
title = {Attention is all you need},
year = {2017},
isbn = {9781510860964},
booktitle = {Proc. 31st Int. Conf.  Neural Inf. Process. Syst. (NeurIPS)},
_old_booktitle = {Proceedings of the 31st International Conference on Neural Information Processing Systems},
pages = {6000-6010},
numpages = {11},
location = {Long Beach, California, USA},
}

@ARTICLE{guoda2024attention,
  author={Tian, Guoda and Pjanić, Dino and Cai, Xuesong and Bernhardsson, Bo and Tufvesson, Fredrik},
  journal={IEEE Trans. Mach. Learn. Commun. Netw.},
  _old_journal={IEEE Transactions on Machine Learning in Communications and Networking}, 
  title={Attention-Aided Outdoor Localization in Commercial {5G} {NR} Systems}, 
  month={November},
  year={2024},
  volume={2},
  number={},
  pages={1678-1692},  
  doi={10.1109/TMLCN.2024.3490496}}

@INPROCEEDINGS{ilayda20255gattention,
  author={Yaman, Ilayda and Tian, Guoda and Pjanić, Dino and Tufvesson, Fredrik and Edfors, Ove and Zhang, Zhengya and Liu, Liang},
  booktitle={Proc. 59th Asilomar Conf. Signals, Syst., Comput.}, 
  _old_booktitle={2025 59th Asilomar Conference on Signals, Systems, and Computers}, 
  title={Adaptive Attention-Based Model for {5G} Radio-Based Outdoor Localization}, 
  month={October},
  year={2025},
  volume={},
  number={},
  pages={192-197},
  address={Grove, CA, USA},
  %doi={10.1109/IEEECONF67917.2025.11443733}
  }

@inproceedings{dosovitskiy2021vit,
  author    = {Alexey Dosovitskiy and
                  Lucas Beyer and
                  Alexander Kolesnikov and
                  Dirk Weissenborn and
                  Xiaohua Zhai and
                  Thomas Unterthiner and
                  Mostafa Dehghani and
                  Matthias Minderer and
                  Georg Heigold and
                  Sylvain Gelly and
                  Jakob Uszkoreit and
                  Neil Houlsby},
  title     = {An Image is Worth 16x16 Words: Transformers for Image Recognition at Scale},
  booktitle = {Proc. 9th Int. Conf. Learn. Represent. (ICLR)},
  _old_booktitle = {Proceedings of the 9th International Conference on Learning Representations (ICLR)},
  location  = {Virtual Event, Austria},
  month     = {May},
  year      = {2021},
  %publisher = {OpenReview.net}, 
}

@ARTICLE{Hyung2025vit,
  author={Cho, Hyung Joon and Ahn, Yongjun and Shim, Byonghyo},
  journal=IEEE_J_VT,
  _old_journal={IEEE Transactions on Vehicular Technology}, 
  title={Transformer-Aided Mobile Positioning for {6G} Ultra-Dense Networks}, 
  month={December},
  year={2025},
  volume={74},
  number={4},
  pages={6839-6843},
  doi={10.1109/TVT.2024.3513439}
  }

@ARTICLE{Gong2024vit,
  author={Gong, Xinrui and Lu, An'an and Liu, Xiaofeng and Fu, Xiao and Gao, Xiqi and Xia, Xiang-Gen},
  journal=IEEE_J_VT,
  _old_journal={IEEE Transactions on Vehicular Technology}, 
  title={Deep Learning Based Fingerprint Positioning for Multi-Cell Massive {MIMO-OFDM} Systems}, 
  month={October},
  year={2024},
  volume={73},
  number={3},
  pages={3832-3849},
  doi={10.1109/TVT.2023.3326825}
  }

@inproceedings{he2016deep,
  title={Deep residual learning for image recognition},
  author={He, Kaiming and Zhang, Xiangyu and Ren, Shaoqing and Sun, Jian},
  booktitle={Proc. IEEE Conf. Comput. Vis. Pattern Recognit. (CVPR)},
  _old_booktitle={Proceedings of the IEEE conference on computer vision and pattern recognition},
  pages={770--778},
  month={June },
  year={2016},
  address={Vegas, NV, USA}
}

@INPROCEEDINGS{lte_ic,
  author={Priyanto, Basuki E. and Kant, Shashi and Rusek, Fredrik and Hu, Sha and Chen, Jianjun and Wugengshi, Chris},
  booktitle={Proc. IEEE 78th Veh. Technol. Conf.}, 
  title={Robust {UE} Receiver with Interference Cancellation in {LTE} Advanced Heterogeneous Network}, 
  year={2013},
  volume={},
  number={},
  pages={},
address={Las Vegas, NV, USA},
  doi={10.1109/VTCFall.2013.6692396}}

@INPROCEEDINGS{rimax_ic,
    author      ={Chen, Junshi and Whiton, Russ and Li, Xuhong and Tufvesson, Fredrik},
    booktitle   ={Proc. EuCNC/6G Summit}, 
    title       ={High-Resolution Channel Sounding and Parameter Estimation in Multi-Site Cellular Networks}, 
    year        ={2023},
    month       ={Jun.},
    volume      ={},
    number      ={},
    pages       ={90-95},
    address     ={Gothenburg, Sweden}, 
}

@ARTICLE{Khalife2019dop,
  author={Khalife, Joe J. and Kassas, Zaher Zak M.},
  _old_journal={IEEE Transactions on Aerospace and Electronic Systems},
journal = IEEE_J_AES,
  title={Optimal Sensor Placement for Dilution of Precision Minimization Via Quadratically Constrained Fractional Programming}, 
  month={Nov.},
  year={2019},
  volume={55},
  number={4},
  pages={2086-2096},
  doi={10.1109/TAES.2018.2879552}}

@ARTICLE{Li2020dop,
  author={Li, Binghao and Zhao, Kai and Shen, Xuesong},
  journal={IEEE Access}, 
  title={Dilution of Precision in Positioning Systems Using Both Angle of Arrival and Time of Arrival Measurements},
  month={Oct.},
  year={2020},
  volume={8},
  number={},
  pages={192506-192516},
  doi={10.1109/ACCESS.2020.3033281}}

@article{langley1999dilution,
  title={Dilution of precision},
  author={Langley, Richard B},
  journal={GPS world},
  volume={10},
  number={5},
  pages={52--59},
  year={1999}
}

@ARTICLE{junshi2026gmf,
  author={Chen, Junshi and Li, Xuhong and Whiton, Russ and Leitinger, Erik and Tufvesson, Fredrik},
  journal={IEEE Open J. Signal Process.},
  _old_journal={IEEE Open Journal of Signal Processing}, 
  title={Robust Localization in Modern Cellular Networks Using Global Map Features}, 
  month={February},
  year={2026},
  volume={7},
  number={},
  pages={356-372},
    doi={10.1109/OJSP.2026.3665385}
  }

@INPROCEEDINGS{wu2021cvt,
  author={Wu, Haiping and Xiao, Bin and Codella, Noel and Liu, Mengchen and Dai, Xiyang and Yuan, Lu and Zhang, Lei},
  booktitle={IEEE Int. Conf. Comput. Vis. (ICCV)},
  _old_booktitle={2021 IEEE/CVF International Conference on Computer Vision (ICCV)}, 
  title={{C}v{T}: Introducing Convolutions to Vision {T}ransformers}, 
  month={October},
  year={2021},
  volume={},
  number={},
  pages={22-31},
  address={Montreal, QC, Canada},
  doi={10.1109/ICCV48922.2021.00009}
  }

@inproceedings{lin2023ConvFormer,
author = {Lin, Xian and Yan, Zengqiang and Deng, Xianbo and Zheng, Chuansheng and Yu, Li},
title = {Conv{F}ormer: Plug-and-Play {CNN}-Style {T}ransformers for Improving Medical Image Segmentation},
month={October},
year = {2023},
isbn = {978-3-031-43900-1},
%publisher = {Springer-Verlag},
address = {Berlin, Heidelberg},
%url = {https://doi.org/10.1007/978-3-031-43901-8_61},
doi = {10.1007/978-3-031-43901-8_61},
booktitle={Proc. 26th Int. Conf. Med. Image Comput. Comput.-Assist. Intervent. ({MICCAI})},
_old_booktitle = {Medical Image Computing and Computer Assisted Intervention – MICCAI 2023: 26th International Conference,  Proceedings},
pages = {642–651},
numpages = {10},
location = {Vancouver, BC, Canada}
}

@INPROCEEDINGS{gu2023conformer,
  author={Gu, Pengfei and Zhang, Yejia and Wang, Chaoli and Chen, Danny Z.},
  booktitle={Proc. IEEE 20th Int. Symp. Biomed. Imag. ({ISBI})}, 
  title={Conv{F}ormer: Combining {CNN} and {T}ransformer for Medical Image Segmentation}, 
  month={April},
  year={2023},
  volume={},
  number={},
  pages={},
  address={Cartagena, Colombia}
  }

@inproceedings{d2021convit,
  title={Con{V}i{T}: Improving vision {T}ransformers with soft convolutional inductive biases},
  author={d’Ascoli, St{\'e}phane and Touvron, Hugo and Leavitt, Matthew L and Morcos, Ari S and Biroli, Giulio and Sagun, Levent},
  booktitle={Proc. Int. Conf. Mach. Learn. ({ICML})},
  pages={2286--2296},
  month={July},
  year={2021},
  %organization={PMLR}
}

@INPROCEEDINGS{guo2022cmt,
  author={Guo, Jianyuan and Han, Kai and Wu, Han and Tang, Yehui and Chen, Xinghao and Wang, Yunhe and Xu, Chang},
  booktitle={Proc. IEEE Conf. Comput. Vis. Pattern Recognit. (CVPR)}, 
  title={{CMT}: Convolutional Neural Networks Meet Vision {T}ransformers}, 
  month={June},
  year={2022},
  volume={},
  number={},
  pages={12165-12175},
  address={New Orleans, LA, USA},
  }

@phdthesis{rimax_richter,
  author  = {A. {Richter}},
  title   = {Estimation of Radio Channel Parameters, Models and Algorithms},
  school  = {Technische Universit{\"a}t Ilmenau},
  year    = {2005},
  address      = {Ilmenau, Germany},
}

@ARTICLE{rui_hrpe_2017,
author={Wang, Rui and Renaudin, Olivier and Bas, C. Umit and Sangodoyin, Seun and Molisch, Andreas F.},
journal=IEEE_J_WCOM,
%journal={IEEE Transactions on Wireless Communications}, 
title={High-Resolution Parameter Estimation for Time-Varying Double Directional {V2V} Channel}, 
month={August},
year={2017}, 
volume={16},
number={11}, 
pages={7264-7275},
doi={10.1109/TWC.2017.2744628},
}

@article{studer2018channel,
  title={Channel Charting: Locating Users Within the Radio Environment Using Channel State Information},
  author={Studer, Christoph and Medjkouh, Sa{\"\i}d and Gonulta{\c{s}}, Emre and Goldstein, Tom and Tirkkonen, Olav},
  journal={IEEE Access},
  volume={6},
  pages={47682--47698},
  month={August},
  year={2018},
  publisher={IEEE},
  doi={10.1109/ACCESS.2018.2866979}
}

@book{thrun2005pr,
author = {Thrun, Sebastian and Burgard, Wolfram and Fox, Dieter},
title = {Probabilistic Robotics},
series = {Intelligent robotics and autonomous agents},
address = {Cambridge, MA, USA},
year = {2005},
isbn = {0262201623},
publisher = {The MIT Press}
}

@misc{NIUSRP,
author = {{National Instruments}},
title = {{URSP-2953 - NI}},
year = {2023},
url = {https://www.ni.com/docs/en-US/bundle/usrp-2953-specs/page/specs.html},
}

@misc{Rubidium_Link,
author = {{Stanford Research Systems}},
title = {Rubidium Frequency Standard - {FS725}},
year = {2020},
url = {https://www.thinksrs.com/products/fs725.html},
}

@misc{oxts,
    author = {{Oxford Technical Solutions Ltd}},
    title = {{RT}3000 v3},
    year = {2020},
    url = {https://www.oxts.com/products/rt3000/},
  }

@INPROCEEDINGS {Hu2018SEN,
author = { Hu, Jie and Shen, Li and Sun, Gang },
booktitle = { Proc. IEEE Conf. Comput. Vis. Pattern Recognit. (CVPR) },
title = {Squeeze-and-Excitation Networks},
year = {2018},
volume = {},
ISSN = {},
pages = {7132-7141},
doi = {10.1109/CVPR.2018.00745},
%publisher = {IEEE Computer Society},
address = {Los Alamitos, CA, USA},
month ={June}
}

@ARTICLE{hu2020SEN,
  author={Hu, Jie and Shen, Li and Albanie, Samuel and Sun, Gang and Wu, Enhua},
  journal=IEEE_J_PAMI,
  %journal={IEEE Transactions on Pattern Analysis and Machine Intelligence}, 
  title={Squeeze-and-Excitation Networks}, 
  month={Aug},
  year={2020},
  volume={42},
  number={8},
  pages={2011-2023},  
  doi={10.1109/TPAMI.2019.2913372}}

@inproceedings{wu2018groupnorm,
author = {Wu, Yuxin and He, Kaiming},
title = {Group Normalization},
month={September},
year = {2018},
isbn = {978-3-030-01260-1},
%publisher = {Springer-Verlag},
%address = {Berlin, Heidelberg},
doi = {10.1007/978-3-030-01261-8_1},
booktitle = {Proc. 15th Eur. Conf. Comput. Vis. (ECCV)},
pages = {3-19},
numpages = {17},
location = {Munich, Germany}
}

@article{hendrycks2016gelu,
  title={Gaussian Error Linear Units ({GELU}s)},
  author={Hendrycks, Dan and Gimpel, Kevin},
  journal={arXiv preprint arXiv:1606.08415},
  year={2016}
}

@ARTICLE{guoda2024fcnn,
  author={Tian, Guoda and Yaman, Ilayda and Sandra, Michiel and Cai, Xuesong and Liu, Liang and Tufvesson, Fredrik},
  journal={IEEE Trans. Mach. Learn. Commun. Netw.}, 
  title={Deep-Learning-Based High-Precision Localization With Massive {MIMO}}, 
  month={November},
  year={2024},
  volume={2},
  number={},
  pages={19-33},
  doi={10.1109/TMLCN.2023.3334712}}

@ARTICLE{Brambilla2024gnss5g,
  author={Brambilla, Mattia and Alghisi, Marianna and Camajori Tedeschini, Bernardo and Fumagalli, Alessandro and Catalin Grec, Florin and Italiano, Lorenzo and Pileggi, Chiara and Biagi, Ludovico and Bianchi, Stefano and Gatti, Andrea and Goia, Alessandro and Nicoli, Monica and Realini, Eugenio},
  %journal={IEEE Open Journal of the Communications Society}, 
  journal=IEEE_OJ_COM,
  title={Integration of {5G} and {GNSS} Technologies for Enhanced Positioning: An Experimental Study}, 
  year={2024},
  month={October},
  volume={5},
  number={},
  pages={7197-7215},  
  doi={10.1109/OJCOMS.2024.3487270}}

@Article{christian2024dmimo,
AUTHOR = {Nelson, Christian and Li, Xuhong and Fedorov, Aleksei and Deutschmann, Benjamin and Tufvesson, Fredrik},
TITLE = {Distributed {MIMO} Measurements for Integrated Communication and Sensing in an Industrial Environment},
JOURNAL = {Sensors},
VOLUME = {24},
YEAR = {2024},
NUMBER = {5},
ARTICLE-NUMBER = {1385},
%URL = {https://www.mdpi.com/1424-8220/24/5/1385},
PubMedID = {38474920},
ISSN = {1424-8220},
DOI = {10.3390/s24051385}
}

@INPROCEEDINGS{Benjamin2024dmimo,
  author={Deutschmann, Benjamin and Nelson, Christian and Henriksson, Mikael and Marti, Gian and Kosasih, Alva and Tervo, Nuutti and Leitinger, Erik and Tufvesson, Fredrik},
  %booktitle={2024 IEEE 25th International Workshop on Signal Processing Advances in Wireless Communications (SPAWC)}, 
  booktitle={Proc. 2024 IEEE 25th Int. Workshop Signal Process. Adv. Wireless Commun. (SPAWC)},
  title={Accurate Direct Positioning in Distributed {MIMO} Using Delay-Doppler Channel Measurements}, 
  year={2024},
  month={October},
  volume={},
  number={},
  pages={606-610},  
  doi={10.1109/SPAWC60668.2024.10694279}}
